\documentclass[preprint,aps,eqsecnum,showpacs,tightenlines]{revtex4}
\usepackage{epsfig}
\newcommand{\be}{\begin{equation}}
\newcommand{\ee}{\end{equation}}
\newcommand{\bea}{\begin{eqnarray}}
\newcommand{\eea}{\end{eqnarray}}

\begin{document}
\preprint{DO-TH 01/07, LA-01-2141, SUSX-TH/01-019}
\title{Dynamics of coupled bosonic systems with applications to preheating}
\author{$^{a}$Daniel Cormier,~ $^{b}$Katrin Heitmann,
and ~$^{c}$Anupam Mazumdar}
\affiliation{$^{a}$ Centre for Theoretical Physics, University of Sussex, Falmer, Brighton,
BN1~9QH, United Kingdom\\
$^{b}$T-8, Theoretical Division, Los Alamos National Laboratory, Los Alamos, New Mexico, 
87545, U.S.A\\
$^{c}$ The Abdus Salam International Centre for Theoretical Physics, 
Strada Costiera, I-10, 34100 Trieste, Italy}
\date{May, 2001}
\begin{abstract}
Coupled, multi-field models of inflation can provide several
attractive features unavailable in the case of a single inflaton
field. These models have a rich dynamical structure resulting from the
interaction of the fields and their associated fluctuations. We
present a formalism to study the nonequilibrium dynamics of
coupled scalar fields. This formalism solves the problem of
renormalizing interacting models in a transparent way using
dimensional regularization. The evolution is generated by a
renormalized effective Lagrangian which incorporates the dynamics of
the mean fields and their associated fluctuations at one-loop
order. We apply our method to two problems of physical interest: (i) a
simple two-field model which exemplifies applications to reheating in
inflation, and (ii) a supersymmetric hybrid inflation model. This
second case is interesting because inflation terminates via a smooth
phase transition which gives rise to a spinodal instability in one of
the fields. We study the evolution of the zero mode of the fields and
the energy density transfer to the fluctuations from the mean fields.
We conclude that back reaction effects can be significant over a wide
parameter range. In particular for the supersymmetric hybrid model we
find that particle production can be suppressed due to these effects.
\end{abstract}
\maketitle

\section{Introduction}

In recent years, the study of nonequilibrium dynamics in quantum field
theory has received much attention in various areas of physics, and
particularly in cosmology.  The work has been driven largely by
inflation \cite{inflation}, the most successful known mechanism for
explaining the large-scale homogeneity and isotropy of the universe
{\em and} the small-scale inhomogeneity and anisotropy of the universe
\cite{cmb}.  With observations for the first time able to directly
test the more detailed predictions of specific inflationary models,
the efforts in understanding inflation and its dynamics have
redoubled.

\vskip7pt

One area of particular interest is the dynamics of multi-field models
of inflation in which the inflaton is coupled to another dynamical
field during inflation. These models can lead to a variety of features
unavailable in the case of a single field.  Such multi-field scenarios
include the well known hybrid inflation models \cite{hybrid}.

\vskip7pt

On top of the dynamics during inflation, the subsequent phase of
energy transfer between the inflaton and other degrees of freedom
leading to the standard picture of Big Bang Cosmology has been the
subject of intense study.  The inflaton may decay through perturbative
processes \cite{albrecht,boyanovsky1} as well as non-perturbative
parametric amplification \cite{preheat,preheat2}.  The latter can lead to
explosive particle production and very efficient reheating of the
universe.

\vskip7pt

Hybrid inflation and reheating models share an important common
thread.  They both involve the coupling of two or more dynamical,
interacting scalar fields (or higher spin fields \cite{higherspin}).
An important aspect of such systems is the possibility of mixing
between the fields. In Ref.~\cite{proc} for example the classical
inflaton decay is investigated for a two field model by solving
the non-linear equations of motions on a grid.
In Ref.~\cite{mixing}, the authors treat the
problem of coupled quantum scalar and fermion fields at the tree
level. Due to the small couplings involved in inflationary cosmology,
such a tree level analysis is useful in a variety of physical
situations.

\vskip7pt

However, hybrid models as well as the dynamics of reheating typically
include processes such as spinodal decomposition \cite{spinodal,spinodal2} and
parametric amplification which require one to go beyond the tree level
by including quantum effects either in a perturbative expansion or by
means of non-perturbative mean field techniques such as the Hartree
approximation or a large-$N$ expansion \cite{boyanovsky1,thesis,LosAl}.

\vskip7pt

Going beyond tree level brings in the issue of renormalization. The
problem of renormalization of time evolution equations in single field
models was understood several years ago. In one of the first papers in
this field, Cooper and Mottola showed in 1987
(Ref. \cite{fred}, that it is possible to find a renormalization
procedure which leads to counter terms independent of time and initial
conditions of the mean field. They used a WKB expansion in order to
extract the divergences of the theory. In a later paper Cooper et
al. also discussed a closely related adiabatic method in order to
renormalize the $\phi^4$-theory in the large N approximation. Also
Boyanovsky and de Vega, Ref. \cite{spinodal}, used a WKB method
in order to renormalize time dependent equations in one-loop order,
later on Boyanovsky et al.  \cite{spinodal2} investigated a $\phi^4$
model in the large-N approximation and the Hartree approximation,
too. In 1996 Baacke et al., Ref. \cite{katrin0}, proposed a slightly
different method in order to extract the divergences of the theory,
which enabled them to use dimensional regularization. In contrast to
the WKB ansatz this method can be extended for coupled
system, which was demonstrated in Ref. \cite{katrin2}. This procedure
will be used also in this paper.  We work in the context of a closed
time path formalism \cite{ctp} appropriate to following the
time-dependent evolution of the system.  In this formalism, the {\em
in}-vacuum plays a predominant role, as quantities are tracked by
their {\em in-in} expectation values (in contrast to the {\em in-out}
formalism of scattering theory).  We construct the {\em in}-vacuum by
diagonalizing the mass matrix of the system at the initial time $t=0$.
However, because of the time-dependent mixing, a system initially
diagonalized in this way will generally not be diagonalized at later
times.

\vskip7pt

One approach to this problem, taken in Ref.~\cite{mixing}, is to
diagonalize the mass matrix at each moment in time through the use of
a time-dependent rotation matrix.  The cost of doing so is the
appearance of time derivatives of the rotation matrix into the kinetic
operators of the theory.  While such a scheme is in principle workable
beyond the tree level, the modified kinetic operators introduce
complications into the extraction of the fluctuation corrections as well
as the divergences that are to be removed via renormalization.

\vskip7pt

We take an alternative approach where the mass matrix is allowed to be
non-diagonal for all times $t>0$ and account for the mixing by
expanding each of the fields in terms of {\em all} of the {\em
in}-state creation and annihilation operators.  The cost of doing so in
an $N$-field system is the need to track $N^2$ complex mode functions
representing the fields instead of the usual $N$.  However, this
allows standard techniques to be used to properly renormalize the
system.  For the two-field systems common in inflationary models, this
effective doubling of the field content adds a relatively minor cost.

\vskip7pt

For simplicity and clarity, we will work in Minkowski space time and in
a one-loop approximation.  Extensions both to
Friedmann-Robertson-Walker spacetimes and to simple non-perturbative
schemes such as the Hartree approximation, while more complicated than
the present analysis, present no fundamental difficulties.  We note
that Minkowski space time is a good approximation in the latter stages
of certain hybrid inflation models, and it will also allow comparison
with much of the original reheating literature
\cite{albrecht,boyanovsky1,preheat} which often neglects the effects
of expansion, allowing us to directly determine the role played by the
mixing of the fields in the dynamics.

\vskip7pt

The outline of the paper is as follows.  We begin by considering the
Lagrangian for $N$ coupled scalar fields and set up our formalism for
the quantization of the system.  This is followed by an outline of the
renormalization procedure.  We then provide a summary of the results for
the two-field case.  We demonstrate the formalism with two examples: a
simple reheating model and a hybrid inflation model motivated by
supersymmetry.

\vskip7pt

In the reheating model we investigate two relevant regimes discussed in
detail in the literature \cite{preheat,preheat2}, viz., the narrow resonance
regime and the broad resonance regime. These different regimes occur
depending on the choice of initial conditions. Usually in these models
the mixing effects of the fields were neglected by choosing a
vanishing initial value for one of the mean fields: We are now able to
treat the full system and to investigate these mixing effects. For
this purpose we concentrate on studying the behavior of the
fluctuation integrals for the different fields and the time-dependent
mixing angle. Depending on the regime, as the mean fields evolve, the
effects of the mixing can be quite different. In the narrow resonance
regime the mixing angle is very small and plays a sub-dominant role,
whereas in the broad resonance regime the mixing effects are very
important. Therefore, we emphasize that neglecting the mixing could
lead to incomplete results.

\vskip7pt

Supersymmetric hybrid models are a special realization of general
hybrid inflationary models (see e.g. Refs. \cite{mar,schmidt}). Based on
a softly broken supersymmetry potential, the special feature of these
models is the occurrence of only one coupling constant, whereas in
nonsupersymmetric hybrid models there are at least two different
couplings. Thus, in the supersymmetric case there is only one natural
frequency of oscillation for the mean fields as long as fluctuations
are neglected. This leads to efficient particle production during the
preheating stage in the early universe. However, we show below that,
by taking into account the fluctuations and investigating the full
mixed system, this feature of supersymmetric hybrid models can be lost
in some regimes. This is because the effective mass corrections for
the two fields are different in these regimes, which leads to a chaotic
trajectory for the renormalized field equations of motion in a phase 
space which mimics the situation of a nonsupersymmetric hybrid model. 
It appears, then, that supersymmetric hybrid models can lose some of 
their attractiveness compared to general hybrid models.


\section{N Fields}

We work with the following Lagrangian for real scalar fields $\Phi_i$
with $i=1 \dots N$:
\begin{equation}
L[\Phi_i] = \sum_{i=1}^N \frac12 \partial_\mu \Phi_i(x) \partial^\mu 
\Phi_i(x) - V[\Phi_i(x)] \; , 
\end{equation}
where the potential is
\begin{equation}
V(\Phi_i) = \sum_{i,j,k,l=1}^N A_i \Phi_i + \frac12 m_{ij} \Phi_i \Phi_j
+ \frac{1}{3!} g_{ijk} \Phi_i \Phi_j \Phi_k 
+ \frac{1}{4!} \lambda_{ijkl} \Phi_i \Phi_j \Phi_k \Phi_l \; .
\label{pot}
\end{equation}
Note that $m_{ij}$, $g_{ijk}$, and $\lambda_{ijkl}$ are symmetric in
each index, but are generally non-diagonal resulting in the mixing
of the different fields.  In what follows, subscripts and superscripts
$i \dots n$ run over the values $1 \dots N$ and we use a convention
in which summation is assumed over repeated lowered indices, but 
not raised indices.  

\vskip7pt

We will expand each field about their expectation values (taken to be
space translation invariant): 
\begin{equation}
\Phi_i(x) = \phi_i(t) + \delta \phi_i(x) \; , \quad 
\phi_i(t) = \langle \Phi_i(x) \rangle \; .
\end{equation}
Expanding the equations of motion and keeping terms to quadratic order
in the  fluctuations yields a one loop approximation.
The equations of motion for the zero modes $\phi_i$ are
determined via the tadpole condition.  We have
\begin{eqnarray}
\ddot{\phi_i} + A_i &+& m_{ij} \phi_j + \frac12 g_{ijk} \phi_j \phi_k
+ \frac16 \lambda_{ijkl} \phi_j \phi_k \phi_l \nonumber \\
&+& \frac12 g_{ijk} \langle \delta \phi_j \delta \phi_k \rangle
+ \frac12 \lambda_{ijkl} \phi_j \langle \delta \phi_k \delta \phi_l \rangle
 = 0 \; .
\label{phiieq}
\end{eqnarray}
To this order, the fluctuations obey the equation
\begin{equation}
\ddot{\delta \phi}_i - \vec{\nabla}^2 \delta \phi_i 
+ {\cal M}_{ij} \delta \phi_j = 0 \; ,
\end{equation}
with the mass matrix
\begin{equation}
{\cal M}_{ij} = m_{ij} + g_{ijk} \phi_k 
+ \frac12 \lambda_{ijkl} \phi_k \phi_l \; .
\label{Mij}
\end{equation}

As indicated in the introduction,
the complication that arises is not the fact that the mass matrix 
(\ref{Mij}) contains mixing between the various fields, rather that 
the mixing changes with time as the $\phi_i$ evolve according to 
(\ref{phiieq}).  This means that if we diagonalize the mass matrix
at one time, it will not generally be diagonal at any other time.

\vskip7pt

Nonetheless, it is most convenient to quantize in terms of a diagonal
system at the initial time $t=0$.  We define the matrix
\begin{equation}
{\cal D}_{ij} = {\cal O}_{ik} {\cal M}_{kl} {\cal O}^T_{lj} \; ,
\label{Dij}
\end{equation}
and the corresponding fluctuation fields
\begin{equation}
X_i = {\cal O}_{ij} \delta \phi_j \; ,
\end{equation}
where ${\cal O}_{ij}$ is an orthogonal rotation matrix.
${\cal D}_{ij}$ is diagonal at the initial time:
\begin{equation}
{\cal D}_{ij} = D^i \delta_{ij} \; ,
\end{equation}
without summation over the raised index $i$.
The $X_i$ obey the equations of motion
\begin{equation}
\ddot{X}_i - \vec{\nabla}^2 X_i + {\cal D}_{ij} X_j = 0 \; .
\end{equation}

We quantize the system by defining a set of creation and annihilation
operators $a^{\dagger}_\alpha(\vec{k})$ and $a_\alpha(\vec{k})$ where
$\alpha = 1 \dots N$ corresponds to the {\em in}-state quanta of
frequency 
\begin{equation}
\omega_\alpha = \sqrt{k^2 + D^\alpha} \; .
\end{equation}
As the mixing changes in time, each of the fields $X_i$ is expanded
in terms of all of the {\em in}-state operators.  We have
\begin{equation}
X_i = \sum_{\alpha=1}^N \int \frac{d^3k}{(2\pi)^3} 
\frac{1}{2\omega_{\alpha 0}} \left[
a_\alpha(\vec{k}) U_i^\alpha(\vec{k},t) e^{i\vec{k} \cdot \vec{x}}
+ a^{\dagger}_\alpha(\vec{k}) U_i^{\alpha*}(\vec{k},t) 
e^{-i\vec{k} \cdot \vec{x}} \right] \; .
\end{equation}
The initial conditions for the $N^2$ complex mode functions are
\begin{equation}
U_i^\alpha(\vec{k},0) = \delta_i^\alpha
\; , \quad 
\dot{U}_i^\alpha(\vec{k},0) = -i\omega_\alpha U_i^\alpha(\vec{k},0)\; .
\end{equation}

It is convenient to define the fluctuation integrals
\begin{equation}
\langle X_i X_j \rangle = \sum_{\alpha=1}^N \int \frac{d^3k}{(2\pi)^3}
\frac{1}{2\omega_{\alpha 0}}
U_i^{\alpha*}(\vec{k},t) U_j^\alpha(\vec{k},t) \; ,
\label{Xijfluct}
\end{equation} 
from which it is straightforward to determine the contributions 
appearing in the zero mode equations (\ref{phiieq}):
\begin{equation}
\langle \delta \phi_i \delta \phi_j \rangle = {\cal O}^T_{ik} {\cal O}^T_{jl} 
\langle X_i X_j \rangle
 \; .
\end{equation}
It will also prove convenient to introduce the rotated couplings
\begin{eqnarray}
G_{ijk} &=& g_{ilm} {\cal O}^T_{lj} {\cal O}^T_{mk} \; , \\
\Lambda_{ijkl} &=& \lambda_{ijmn} {\cal O}^T_{mk} {\cal O}^T_{nl} \; ,
\end{eqnarray}
which allows us to write the zero mode equations as
\begin{eqnarray}
\ddot{\phi}_i + A_i &+& m_{ij} \phi_j + \frac12 g_{ijk} \phi_j \phi_k
+ \frac16 \lambda_{ijkl} \phi_j \phi_k \phi_l \nonumber \\
&+& \frac12 G_{ijk} \langle X_j X_k \rangle
+ \frac12 \Lambda_{ijkl} \phi_j \langle X_k X_l \rangle
 = 0 \; ,
\label{phiieqX}
\end{eqnarray}
while the mode functions obey the equations
\begin{equation}
\ddot{U}_i^\alpha(\vec{k},t) + \left(k^2 + {\cal D}_{ij}\right)
U_j^\alpha(\vec{k},t) = 0 \; .
\label{mod}
\end{equation}

In addition to the equations of motion, it is useful to have an
expression for the energy density of the system.  This is particularly
true when one completes numerical simulations of the system, since
energy conservation is a powerful check of the accuracy of the 
simulations.  After once again decomposing the fields into their
expectation values and fluctuations, the energy density to one
loop order is
\begin{eqnarray}
\label{energy1}
{\cal E} &=&  \frac12 \dot{\phi}^2_i 
+  A_i \phi_i + \frac12 m_{ij} \phi_i \phi_j
+ \frac{1}{3!} g_{ijk} \phi_i \phi_j \phi_k 
+ \frac{1}{4!} \lambda_{ijkl} \phi_i \phi_j \phi_k \phi_l \nonumber \\
&+&  \frac12 \left\langle \dot{X}^2_i \right\rangle 
+  \frac12 \left\langle 
\left(\vec{\nabla} X_i \right)^2 \right\rangle
+ \frac12 {\cal D}_{ij} \langle X_i X_j \rangle \; ,
\end{eqnarray}
where we've defined the integrals
\begin{eqnarray}
\left\langle \dot{X}_i \right\rangle &=& 
\sum_{\alpha=1}^N \int \frac{d^3k}{(2\pi)^3}
\frac{1}{2\omega_{\alpha 0}}
|\dot{U}_i^{\alpha}(\vec{k},t)|^2  \; ,
\label{Xidotfluct} \\
\left\langle \left(\vec{\nabla} X_i \right)^2 \right\rangle &=&
\sum_{\alpha=1}^N \int \frac{d^3k}{(2\pi)^3}
\frac{k^2}{2\omega_{\alpha 0}}
|{U}_i^{\alpha}(\vec{k},t)|^2  \; .
\label{Xigradfluct}
\end{eqnarray}


\subsection{Divergence structure and renormalization}
\label{renorm}

The mode integrals in the equation of motion defined by
Eq.~(\ref{Xijfluct}) and in the energy density defined by
Eqs. (\ref{Xidotfluct}), (\ref{Xigradfluct}) are divergent and have t
be regulated, allowing for a renormalization of the theory.  We
require a method of extracting the divergent terms appearing in the
mode integrals, a nontrivial task, since the mode equations vary in
time and they are coupled. Our aim is now, to find counter terms, which
are independent of the initial value of the mean fields in order to
formulate a finite theory. The correct choice of the initial condition
for the fluctuations guarantees that the theory is renormalizable.
One way to extract the divergences of the mode integrals is due to a
WKB method which allows for a high momentum expansion of the mode
functions.  However, when the fields are coupled, as in the present
case, the usual formulation of the WKB expansion runs into
difficulties which are yet to be resolved.


\vskip7pt

An alternative method has been developed \cite{katrin2,katrin0,katrin1} which
relies on a formal perturbative expansion in the effective masses and
time derivatives of the masses of the fields.  As such, it results in
a series expansion of the mode functions in powers of $m/\omega$ and
$\dot{m}/\omega^2$, etc.  The first few terms in the series include
the divergent parts of the integrals that are to be removed via
renormalization.

\vskip7pt

We begin by introducing the following ansatz for the mode functions:
\begin{eqnarray}
\label{ansatz0}
U^{\alpha}_{j}= e^{-i\omega_{\alpha 0}t}
\left( \delta^{\alpha}_{j}
+f^{\alpha}_{j}
\right )\,.
\end{eqnarray}
The first term on the right hand side anticipates a quadratic divergence
in the quantities $\langle X_i^2 \rangle$. 
We define the following potential
\begin{eqnarray}
\label{ansatz1}
V^{\alpha}_{ij}(t) = {\cal D}_{ij}(t) - D^\alpha \delta_{ij} \,.
\end{eqnarray}
The equations of motion for 
the mode functions Eqs.~(\ref{mod}) can be written in a suggestive form with 
the help of Eqs.~(\ref{ansatz0},\ref{ansatz1})
\begin{eqnarray}
\ddot f^{\alpha}_{j}- i2\omega_{\alpha 0}\dot f^{\alpha}_{j}=-
\sum_{l=1,2} V^{\alpha}_{jl}\left(\delta^{\alpha}_{l}+f^{\alpha}_{l}\right)\,.
\end{eqnarray}
The terms on the right hand side of this expression are treated
as perturbations to write the $f's$ order by order
in $V$, with the initial conditions $f^{i}_{j}(0)=\dot f^{i}_{j}(0)=0$. 
To first order in $V$, we have the equations of motion:
\begin{eqnarray}
\ddot f^{\alpha(1)}_{j}-2i\omega_{\alpha 0}\dot f^{\alpha(1)}_{j}
 = -V^{\alpha}_{j\alpha}\,.
\label{zerofreq}
\end{eqnarray}
The corresponding integral solutions for the real part of the $f$'s are:
\begin{eqnarray}
2\Re f^{\alpha (1)}_{j}=-\frac{V^{\alpha}_{\alpha j}}{2\omega_{\alpha 0}}
+\int^{t}_{0}dt^{\prime}
\frac{\dot V^{\alpha}_{\alpha j}(t^{\prime})}{2\omega_{\alpha 0}}
\cos (2\omega_{\alpha 0}t^{\prime})\,,
\end{eqnarray} 
while the imaginary part is of order $1/\omega^3$ and does not 
contribute to the divergences \cite{katrin2}.

\vskip7pt

Using these results, we find quadratic and logarithmic divergences:
\begin{equation}
\langle X_i X_j \rangle_{\mbox{div}} = \int \frac{d^3k}{(2\pi)^3}\left( 
\frac{1}{2\omega_{j}} \delta_{ij} - \frac{1}{4\omega^3_j} V^j_{ij}
\right) \; , 
\label{XiXjdiv}
\end{equation}
which must be removed via some renormalization procedure while also 
providing finite corrections to the parameters of the theory.  

\vskip7pt

To make the renormalization scheme explicit, we adopt dimensional
regularization.  We define the following divergent integrals
\begin{eqnarray}
\int \frac{d^3k}{(2\pi)^3} \frac{1}{2\sqrt{k^2+\mu^2}} &=&
-\mu^2 I_{-3}(\mu) - \frac{\mu^2}{16 \pi^2} \; , \\
\int \frac{d^3k}{(2\pi)^3} \frac{1}{4\left(k^2+\mu^2\right)^{3/2}}
&=& I_{-3}(\mu) \; ,
\end{eqnarray}
where $\mu$ is an arbitrary renormalization point and $I_{-3}$ carries
the infinite contributions.  
In dimensional regularization $I_{-3}(\mu)$ is given by
\begin{equation}
I_{-3}(\mu)=\frac{1}{16\pi^2}\left\{\frac 2 \epsilon +\ln\frac{4\pi\mu^2}{m^2}
-\gamma\right\}\; .
\end{equation}
The infinite part of $\langle X_i X_j \rangle$
is found to be simply 
\begin{equation}
\langle X_i X_j \rangle_{\mbox{infinite}} = -{\cal D}_{ij} I_{-3}(\mu)
\; .
\end{equation}

This leads to mass and coupling constant counter terms of the
following form
\begin{eqnarray}
\label{delA}
\delta A_i &=& \frac12 I_{-3}(\mu) g_{ijk} m_{jk} \; , \\
\delta m_{ij} &=& \frac12 I_{-3}(\mu) \left[g_{ikm} g_{kmj}
+ \lambda_{ijkl} m_{kl} \right] \; , \\ 
\delta g_{ijk} &=& \frac32 I_{-3}(\mu) g_{ilm} \lambda_{lmjk} \; , \\
\label{dell}
\delta \lambda_{ijkl} &=& \frac32 I_{-3}(\mu) \lambda_{ijmn}
\lambda_{mnkl} \; . 
\end{eqnarray}
It is important to notice, that these counterterms are independent of
the initial conditions of the mean fields $\phi_i$.

In addition to these counterterms, there are finite 
corrections of the parameters coming from the finite parts of
the integrals (\ref{XiXjdiv}):
\begin{equation}
\langle X_i X_j \rangle_{\mbox{div,finite}} = -\frac{1}{16\pi^2}
D^j \delta_{ij} - \frac{1}{16\pi^2} {\cal D}_{ij} 
\ln \frac{D^j}{\mu^2} \; .
\end{equation}
From this, we extract the following finite contributions to the
couplings and mass:
\begin{eqnarray}
\label{delAf}
\Delta A_i &=& - \frac{1}{32\pi^2} \left[ G_{ijj}D^j + 
g_{ijk} m_{jk} \ln \frac{D^k}{\mu^2} \right] \; , \\
\Delta m_{ij} &=& - \frac{1}{32\pi^2} \left[ \Lambda_{ijkk} D^k
+ \left(g_{ikl}g_{klj} + \lambda_{ijkl} m_{kl}\right) 
\ln \frac{D^k}{\mu^2} \right] \; , \\
\Delta g_{ijk} &=& - \frac{1}{8\pi^2} g_{ilm}\lambda_{lmjk} 
\ln \frac{D^m}{\mu^2} \; , \\
\label{dellf}
\Delta \lambda_{ijkl} &=& - \frac{3}{32\pi^2} \lambda_{ijmn}
\lambda_{mnkl} \ln \frac{D^m}{\mu^2} \; .
\end{eqnarray}

These finite corrections are also contributing to the energy density.
In addition we find a finite part due to the cosmological constant
renormalization.

\vskip7pt

The full, finite equations of motion become
\begin{eqnarray}
&&\ddot{\phi}_i + A_i + \Delta A_i + \left(m_{ij} + \Delta m_{ij} \right) 
\phi_j + \frac12 \left(g_{ijk} + \Delta g_{ijk}\right) \phi_j \phi_k
+ \frac16 \left(\lambda_{ijkl} + \Delta \lambda_{ijkl} \right) 
\phi_j \phi_k \phi_l \nonumber \\
&&+ \frac12 \left(G_{ikl} + \Lambda_{ijkl} \phi_j\right) 
\sum_{\alpha = 1}^{N} \int \frac{d^3k}{(2\pi)^3} \frac{1}{2\omega_{\alpha 0}} 
\left[ f^{\alpha *}_k f^{\alpha}_l + \delta^{\alpha}_k f^{\alpha}_l
+ \delta^{\alpha}_l f^{\alpha *}_k + \frac{1}{2\omega^2_{\alpha 0}} 
\delta^{\alpha}_l V^{\alpha}_{k \alpha} \right]
= 0 \; .
\label{phiieqXren}
\end{eqnarray}


\subsection{Two Fields}

The two-field case is often encountered, and
the physical applications we present in the next section are both in
this category.  It is therefore worthwhile to pause to look 
at a few details of such systems.

\vskip7pt

We begin with a system of two real scalar fields $\Phi$ and $X$
\begin{eqnarray}
\label{lag}
{\cal L}=\frac{1}{2}(\partial_{\mu}\Phi)^2
+\frac{1}{2}(\partial_{\mu}X)^2
-V(\Phi,X)\,,
\end{eqnarray}
with the potential
\begin{equation}
V(\Phi,X) = \frac12 m^2_\phi \Phi^2 + \frac12 m^2_\chi X^2
+ \frac{\lambda}{4!} \Phi^4 + \frac{\kappa}{4!} X^4 
+ \frac{g^2}{4} \Phi^2 X^2 \; .
\end{equation}
This Lagrangian has the same form as that studied in the preceding 
section with the identifications
\begin{eqnarray}\label{ident}
\Phi_1 &\equiv& \Phi \; , \quad \Phi_2 \equiv X \; , \nonumber \\
m_{11} &\equiv& m^2_\phi \; , \quad m_{22} \equiv m^2_\chi \; , 
\quad m_{12} = m_{21} = 0 \; , \nonumber \\
\lambda_{1111} &\equiv& \lambda \; , \quad \lambda_{2222} \equiv \kappa
\; , \quad \lambda_{1122} \equiv g^2  \; , \quad
\lambda_{1112} = \lambda_{1222} = 0\; , \nonumber \\
A_i &=& 0 \; , \quad g_{ijk} = 0  \;  .
\end{eqnarray}
The remaining components of $\lambda_{ijkl}$ are determined by the
fact that it is symmetric in each of its indices.

\vskip7pt

The mass matrix ${\cal M}$ is 
\begin{eqnarray}
{\cal M} = \left( 
\begin{tabular}{cc}
$m_\phi^2 + \lambda \phi^2/2 + g^2 \chi^2/2$ &
$g^2 \phi \chi$ \\
$g^2 \phi \chi$ &
$m_\chi^2 + \kappa 
\chi^2/2 + g^2 \phi^2/2$
\end{tabular}
\right)\;.
\label{M_ij}
\end{eqnarray}

For two fields, the orthogonal rotation matrix can be written 
in terms of a single mixing angle $\theta$.  The matrix has the 
form
\begin{eqnarray}
{\cal O} =
\left( 
\begin{tabular}{cc}
$\cos \theta$ &
$\sin \theta$ \\
$-\sin \theta$ &
$\cos \theta$
\end{tabular}
\right)\;,
\end{eqnarray}
where the mixing angle is determined by the $t=0$ mass matrix 
${\cal M}$, Eq.~(\ref{M_ij}), through the relation
\begin{equation}
\tan \theta = \frac{1}{2{\cal M}_{12}(0)}\left[ {\cal M}_{22}(0) 
- {\cal M}_{11}(0) + \sqrt{\left({\cal M}_{22}(0) 
- {\cal M}_{11}(0) \right)^2 + 4 {\cal M}^2_{12}(0)}
\right]  \; .
\label{thetaeqn}
\end{equation}

The eigenvalues of ${\cal O}$ are the diagonal elements of the 
matrix ${\cal D}$, Eq.~(\ref{Dij}), at the initial time:
\begin{eqnarray}
{\cal D}(0) = 
\left( 
\begin{tabular}{cc}
$D^1$ &
$0$ \\
$0$ &
$D^2$
\end{tabular}
\right)\;,
\end{eqnarray}
with the values
\begin{eqnarray}
D^1 &=& \frac12 \left[ {\cal M}_{11}(0) + {\cal M}_{22}(0)
+ \sqrt{\left({\cal M}_{22}(0) 
- {\cal M}_{11}(0) \right)^2 + 4 {\cal M}^2_{12}(0)} \right]
\; , \\
D^2 &=& \frac12 \left[ {\cal M}_{11}(0) + {\cal M}_{22}(0)
- \sqrt{\left({\cal M}_{22}(0) 
- {\cal M}_{11}(0) \right)^2 + 4 {\cal M}^2_{12}(0)} \right]
\; .
\end{eqnarray}

For general times, the mass matrix for the fields $X_1$ and $X_2$,
writing $c_{\theta}=\cos \theta$ and $s_{\theta}=\sin \theta$, is
\begin{eqnarray}
{\cal D}(t) = 
\left( 
\begin{tabular}{cc}
$c_{\theta}^2{\cal M}_{11}
+2 s_{\theta} c_{\theta} {\cal M}_{12}+ s_{\theta}^2 {\cal M}_{22}$ &
$s_{\theta} c_{\theta} {\cal M}_{22}-s_{\theta} c_{\theta}{\cal M}_{11}
+(c_{\theta}^2 - s_{\theta}^2){\cal M}_{12}$ \\
$s_{\theta} c_{\theta} {\cal M}_{22}-s_{\theta} c_{\theta}{\cal M}_{11}
+(c_{\theta}^2 - s_{\theta}^2){\cal M}_{12}$ &
$-2c_{\theta}^2{\cal M}_{22}
-2 s_{\theta} c_{\theta} {\cal M}_{12}+ s_{\theta}^2 {\cal M}_{11}$
\end{tabular}
\right)
\end{eqnarray}

The zero mode equations, before renormalization, read
\begin{eqnarray}
\label{0mod0}
\ddot \phi + m^2_\phi \phi +\frac{\lambda}{6}\phi^3 + 
\frac{g^2}{2}\phi \chi^2 + \sum_{ij}Q_{ij}(t)\langle
X_i X_j \rangle =0 \,, \\
\label{0mod1}
\ddot \chi + m^2_\chi \chi +\frac{\kappa}{6}\chi^3 + 
\frac{g^2}{2}\chi \phi^2 + \sum_{ij}R_{ij}(t)\langle
X_i X_j \rangle =0 \,,
\end{eqnarray}
where
\begin{eqnarray}
Q_{ij}&=& \Lambda_{1kij}\phi_k \nonumber \\ &=& \left( 
\begin{tabular}{cc} $\frac{\lambda}{2}c_{\theta}^2 \phi +\frac{g^2}{2} 
s_{\theta}^2 \phi +g^2 s_{\theta} c_{\theta} \chi$ &
$-\frac{\lambda}{2} s_{\theta} c_{\theta} \phi+\frac{g^2}{2}
s_{\theta} c_{\theta} \phi+\frac{g^2}{2}(c_{\theta}^2 
- s_{\theta}^2)\chi$ \\
$-\frac{\lambda}{2} s_{\theta} c_{\theta} \phi+\frac{g^2}{2}
s_{\theta} c_{\theta} \phi+\frac{g^2}{2}(c_{\theta}^2 
- s_{\theta}^2)\chi$ &
$\frac{\lambda}{2} s_{\theta}^2 \phi+\frac{g^2}{2}c_{\theta}^2\phi
-g^2 s_{\theta} c_{\theta} \chi$ 
\end{tabular} \right) \,,
\nonumber \\
R_{ij}&=& \Lambda_{2kij}\phi_k \nonumber \\ &=& \left( 
\begin{tabular}{cc} $\frac{\kappa}{2}s_{\theta}^2 \chi +
\frac{g^2}{2} c_{\theta}^2 \chi +g^2 s_{\theta} c_{\theta} \phi$ &
$\frac{\kappa}{2} s_{\theta} c_{\theta} \chi-\frac{g^2}{2}
s_{\theta} c_{\theta} \chi+\frac{g^2}{2}(c_{\theta}^2 
- s_{\theta}^2)\phi$ \\
$\frac{\kappa}{2} s_{\theta} c_{\theta} \chi-\frac{g^2}{2}
s_{\theta} c_{\theta} \chi+\frac{g^2}{2}(c_{\theta}^2 
- s_{\theta}^2)\phi$
& $\frac{\kappa}{2}c_{\theta}^2\chi+\frac{g^2}{2} s_{\theta}^2 \chi
-g^2 s_{\theta} c_{\theta} \phi$  \,.
\end{tabular} \right) \,. 
\end{eqnarray}

The total energy density 
of the system including the fluctuations can be 
expressed as
\begin{eqnarray}
\label{energy0}
{\cal E}&=&\frac{1}{2}\dot\phi^2+ \frac{1}{2}\dot\chi^2+\frac{1}{2}m_{\phi}^2\phi^2
+\frac{1}{2}m_{\chi}^2\chi^2+\frac{\lambda}{4!}\phi^4+\frac{\kappa}{4!}\chi^4+
\frac{g^2}{4}\phi^2\chi^2\nonumber\\
&+&  \frac12 \left\langle \dot{X}^2_i \right\rangle 
+  \frac12 \left\langle 
\left(\vec{\nabla} X_i \right)^2 \right\rangle
+ \frac12 {\cal D}_{ij} \langle X_i X_j \rangle \; .
\end{eqnarray}

Now we have to formulate finite equations of motion and a finite
energy density. We adopt the renormalization procedure of section
\ref{renorm} for the $N$ field case. By using the identifications 
(\ref{ident}) it is to derive the appropriate counterterms for the two
field case from Eqs. (\ref{delA})-(\ref{dell}). We find in particular
\begin{eqnarray} 
\label{counter}
&&\delta m_{\phi}^2=\frac{1}{2}(\lambda
m_{\phi}^2+g^2m_{\chi}^2)I_{-3}(\mu) \,, \quad \delta
m_{\chi}^2=\frac{1}{2}(g^2m_{\phi}^2+\kappa m_{\chi}^2)I_{-3}(\mu)\,, \\
&&\delta \lambda=\frac{3}{2}(\lambda^2+g^4)I_{-3}(\mu)\,, \quad \delta
g^2=\frac{g^2}{2}(\lambda +\kappa+4g^2)I_{-3}(\mu)\,, \quad \delta
\kappa =\frac{3}{2}(g^4+\kappa^2)I_{-3}(\mu)\,.\nonumber
\end{eqnarray}

Of course we get also similar to Eqs. (\ref{delAf})-(\ref{dellf}) in
the $N$ field case finite corrections to the masses and couplings of
the form
\begin{eqnarray}
\label{fin0}
\Delta m^2_{\phi}&=&-\frac{1}{32 \pi^2 }\left\{D^1
(\lambda c_{\theta}^2+g^2 s_{\theta}^2)
+D^2(\lambda s_{\theta}^2+g^2 c_{\theta}^2)+g^2m_{\chi}^2 L_1+
\lambda m_{\phi}^2 L_2\right\}
\,, \\
\label{fin1}
\Delta m^2_{\chi}&=&-\frac{1}{32 \pi^2 }\left \{D^1
(\kappa s_{\theta}^2+g^2c_{\theta}^2)
+D^2(g^2 s_{\theta}^2+\kappa c_{\theta}^2)+g^2m_{\phi}L_2
+\kappa m_{\chi}^2L_1\right \} \,, \\
\label{fin2}
\Delta \lambda &=&-\frac{1}{32 \pi^2 }\left \{\lambda^2 L_2 +g^4 L_1
\right \}\,, \\
\label{fin3}
\Delta g^2 &=&-\frac{3}{32\pi^2}\left \{g^4 \left(L_1 + L_2 \right)
+\frac{1}{2}g^2\lambda L_2
+\frac{1}{2}g^2 \kappa L_1
\right \}\,, \\
\label{fin4}
\Delta \kappa &=&-\frac{1}{32 \pi^2 }\left \{g^4L_2+\kappa^2L_1 \right \}\,,
\end{eqnarray}
where
\begin{eqnarray}
L_1 = s_{\theta}^2 \ln \frac{D^1}{\mu^2}
+c_{\theta}^2\ln \frac{D^2}{\mu^2}\,, \quad \quad
L_2 = s_{\theta}^2 \ln \frac{D^2}{\mu^2}
+c_{\theta}^2\ln \frac{D^1}{\mu^2} \,.
\end{eqnarray}
As a result of these finite corrections Eqs.~(\ref{fin0}-\ref{fin4}), 
the total
Lagrangian Eq.~(\ref{lag}) is also modified. This is exactly the renormalized 
Lagrangian which we needed. We also find an 
additional finite contribution to the 
classical Lagrangian given by
\begin{eqnarray}
\Delta{\cal L}=-\frac{g^2 s_{\theta} c_{\theta} \phi \chi}{64\pi^2}\left\{4(D^1-D^2)
+\ln \frac{D^1}{D^2}\left[ (\lambda+g^2)\chi^2+(\kappa+g^2)\phi^2+2
(m_{\chi}^2+m_{\phi}^2)\right]\right \}\,, \nonumber \\
\end{eqnarray}
and, the final zero mode equations for $\phi$ and $\chi$ are 
given by
\begin{eqnarray}
\label{0mod5}
&&\ddot \phi + ( m_{\phi}^2 +\Delta m_{\phi}^2)\phi + \frac{1}{6}(\lambda +\Delta
\lambda)\phi^3+\frac{1}{2}(g^2+\Delta g^2)\phi \chi^2 +\frac{\partial \Delta 
{\cal L}}{\partial \phi}\, \nonumber \\
&&+ \left \{ \sum_{l\alpha}Q_{ll}(t)
\int \frac{d^3k}{(2\pi)^3}\frac{1}{2\omega_{\alpha 0}}\left(2\delta^{\alpha}_{l}
\Re f^{l}_{l}+f^{\alpha}_{l}f^{\alpha \ast}_{l}+\frac{V_{ll}}
{4\omega_{\alpha 0}^3}\right)\, \right. \nonumber \\ &&
\left.
+\sum_{\alpha \neq l}Q_{l\alpha}\int\frac{d^3k}{(2\pi)^3}
\frac{1}{2\omega_{\alpha 0}}
\left( \Re f^{\alpha}_{l}+\Re f^{\alpha}_{\alpha}
\Re f^{\alpha}_{l}+\Im f^{\alpha}_{\alpha}\Im
f^{\alpha}_{l} +\frac{V_{l \alpha}}{2\omega_{\alpha 0}^2}\right)\right \}=0\,,
\end{eqnarray}
and,
\begin{eqnarray}
\label{0mod6}
&&\ddot \chi + ( m_{\phi}^2 +\Delta m_{\phi}^2)\chi + \frac{1}{6}(\kappa +\Delta
\kappa)\chi^3+\frac{1}{2}(g^2+\Delta g^2)\phi \chi^2 +\frac{\partial \Delta 
{\cal L}}{\partial \chi}\, \nonumber \\
&&+ \left \{ \sum_{l\alpha}R_{ll}(t)
\int \frac{d^3k}{(2\pi)^3}\frac{1}{2\omega_{\alpha 0}}\left(2\delta^{\alpha}_{l}
\Re f^{l}_{l}+f^{\alpha}_{l}f^{\alpha \ast}_{l}+\frac{V_{ll}}
{4\omega_{\alpha 0}^3}\right)\, \right. \nonumber \\ &&
\left.
+\sum_{\alpha \neq l}Q_{l\alpha}\int\frac{d^3k}{(2\pi)^3}\frac{1}{2\omega_{\alpha 0}}
\left( \Re f^{\alpha}_{l}+\Re f^{\alpha}_{\alpha}
\Re f^{\alpha}_{l}+\Im f^{\alpha}_{\alpha}\Im
f^{\alpha}_{l} +\frac{V_{l \alpha}}{2\omega_{\alpha 0}^2}\right)\right \}=0\,.
\end{eqnarray}


After writing down the finite zero mode equations of motion we also
have to renormalize the energy density. Again by using the ansatz
(\ref{ansatz0}) we can extract the divergent terms of the fluctuation
integrals in Eq. (\ref{energy0}). In addition to the quadratic and
logarithmic divergences we find a quartic divergence. This leads to a
counter term which acts as a cosmological constant and has the form
\begin{eqnarray}
\delta \Lambda =\frac{1}{4}(m_{\chi}^4+m_{\phi}^4)I_{-3}(m)\,.
\end{eqnarray} 
Altogether the divergent part of the energy density reads:
\begin{eqnarray}
\label{endiv}
{\cal E}_{\rm div}&=&-\frac{I_{-3}(m)}{4}\left
\{(g^2m_{\chi}^2+\lambda m_{\phi}^2)\phi^2 +(g^2m_{\phi}^2\kappa
m_{\chi}^2)\chi^2+\frac{1}{2}g^2(\lambda+\kappa+4g^2)\phi^2
\chi^2\right. \nonumber \\ && \left.
+\frac{1}{4}(\lambda^2+g^4)\phi^4+\frac{1}{4}(\kappa^2+g^4)\chi^4+
m_{\chi}^4+m_{\phi}^4\right \}\,.
\end{eqnarray}
This expression leads of course to the same counter terms we found for
the equations of motion, and therefore also to the same finite
corrections to couplings and masses. Therefore it is straightforward
to formulate a finite energy expression.

Now, we are in a position to discuss the physical applications of our problem. 
This we shall do in the next section, but first, we introduce one more
quantity that is convenient in discussing the degree to which the mixing
plays a role in the dynamics.

\subsection{Time-dependent mixing angle}

To better understand how the system evolves in time, it is useful to
have a measure of how much the fields $\phi$ and $\chi$ mix at each
moment, and how this mixing evolves with the system.  To provide us
with a measure of the mixing, we introduce a time-dependent mixing
angle $\Theta(t)$, which is defined in terms of the time-dependent
mass matrix ${\cal M}(t)$, Eq.~(\ref{M_ij}):
\begin{equation}
\tan \Theta(t) = \frac{1}{2{\cal M}_{12}(t)}\left[ {\cal M}_{22}(t) 
- {\cal M}_{11}(t) + \sqrt{\left({\cal M}_{22}(t) 
- {\cal M}_{11}(t) \right)^2 + 4 {\cal M}^2_{12}(t)}
\right]  \; .
\label{thetateqn}
\end{equation}

\section{Physical applications}
After setting up the technical framework, we are now in a position to
investigate some relevant cosmological multi-field models for
inflation. We begin our analysis with a simple two-field model often
used for studying the phase of parametric amplification. (This phase
occurs just after the completion of inflation in chaotic inflationary
models \cite{preheat,preheat2}.) This model provides a useful context to
analyze the effects due to field mixing. Next we turn our attention to
a supersymmetric hybrid inflationary model, which is of particular
interest in cosmology. As discussed in the literature (see for example
Ref. \cite{mar}) particle production (and hence reheating) in these
models is much more efficient compared to the nonsupersymmetric hybrid
models. Until now the mixing effects in these models have not been
treated fully including back reaction effects of the quantum
fluctuations in the mean field approximation. This approximation does
not take rescattering processes into account and therefore we can not
address the problem of thermalization.

\subsection{Reheating}

The reheating phase in chaotic inflationary models is characterized by
two different regimes, which depend on the chosen initial conditions:
the first is the narrow resonance regime, while the second is the case
of broad resonance.  In order to investigate these regimes we examine
two different parameter sets, where only the initial values for the
zero modes are varied. We find significant differences in the behavior
of the fluctuation integrals as well as the mixing angle in the two
regimes.

\vskip7pt

To make the analysis as simple as possible, we set $\lambda = \kappa =
0$ as well as $m_\chi = 0$.  For the remaining parameters, we set
$m_\phi = 1$, which just acts as a unit of mass, and $g^2 = 0.01$.
With these parameters, the case usually studied in the literature,
$\chi(0) = 0$, does not introduce any mixing between the fields since
the off-diagonal elements of the $\phi$-$\chi$ mass matrix are
proportional to $\chi$. However, taking $\chi(0) = 0$ may not always
be the case. For instance, $\chi$ field could take a large vacuum
expectation value during inflation, provided $\chi$ is treated as a
field other than the inflaton. For the purpose of illustration we may
consider a non-zero initial condition for $\chi$ which is of order its
effective mass $g\phi$ at the beginning of the preheating stage and
examine the consequences. The initial condition for $\phi$ in the
narrow resonance regime is fixed by the condition
$g^2\phi(0)/4m_\phi^2=0.1$ (remember, $g$ is fixed to be $g=0.001$)
and for the broad resonance regime by $g^2\phi(0)/4m_{\phi}^2=100$.
If we take $\phi(0)$ to be approximately the Planck scale as appropriate
to the end of inflation, this would correspond to $m_{\phi} \sim 10^{17} GeV$ 
and $m_{\phi} \sim 10^{14} GeV$ for the two respective cases.  Note
that these parameters are chosen to depict the phenomena of interest
during a time scale that can be accurately simulated.  The results are, in
any case, representative of two field mixing in the narrow and broad
resonance regimes regardless of the precise parameter values in any
particular model.

\noindent
\parbox{14.8cm}{
\parbox[t]{8cm}
{\begin{center}
\mbox{\epsfxsize=7.5cm\epsfbox{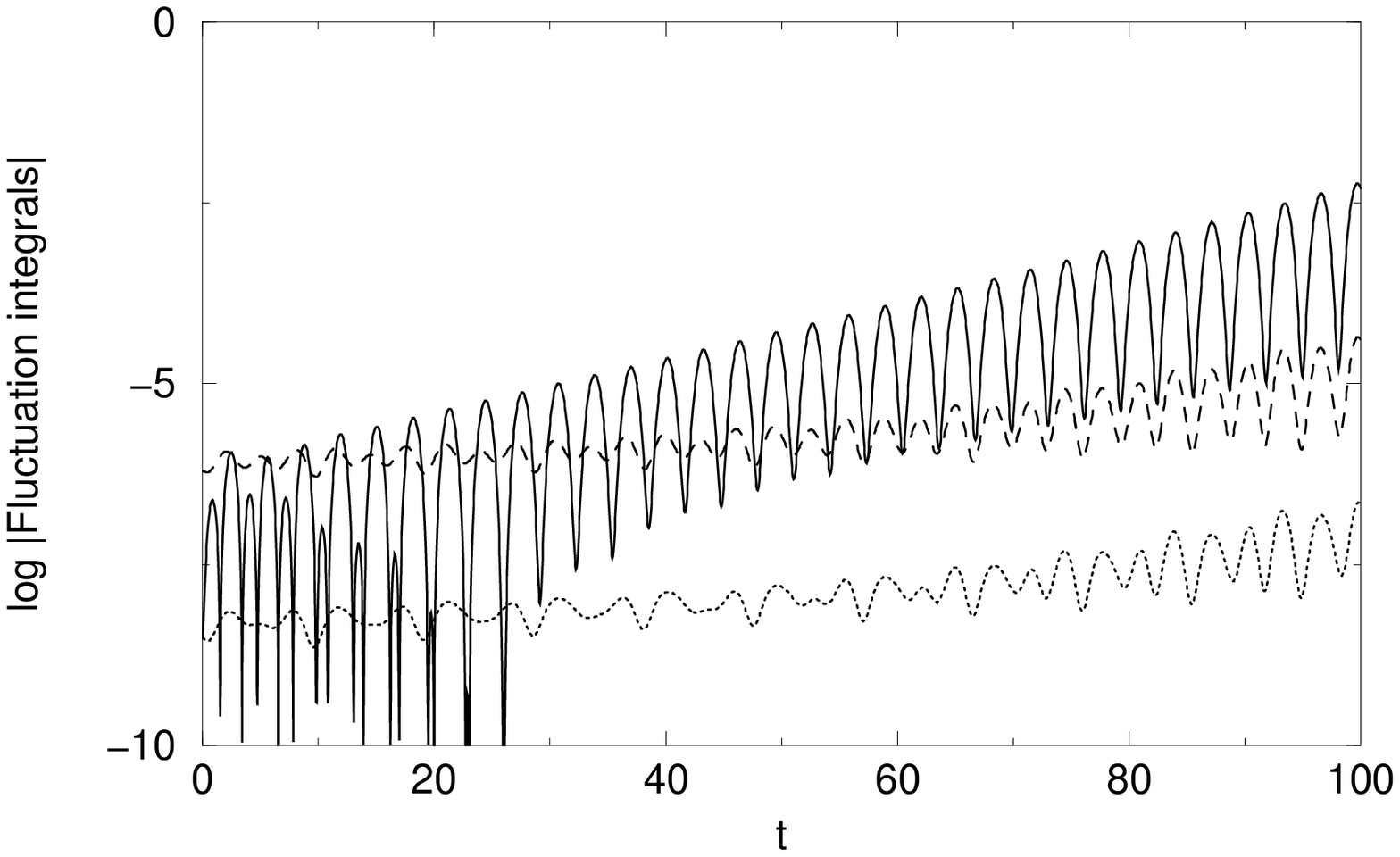}}
\end{center}}
\hspace{.0cm}\parbox[t]{8cm}
{\begin{center}
\mbox{\epsfxsize=7.5cm\epsfbox{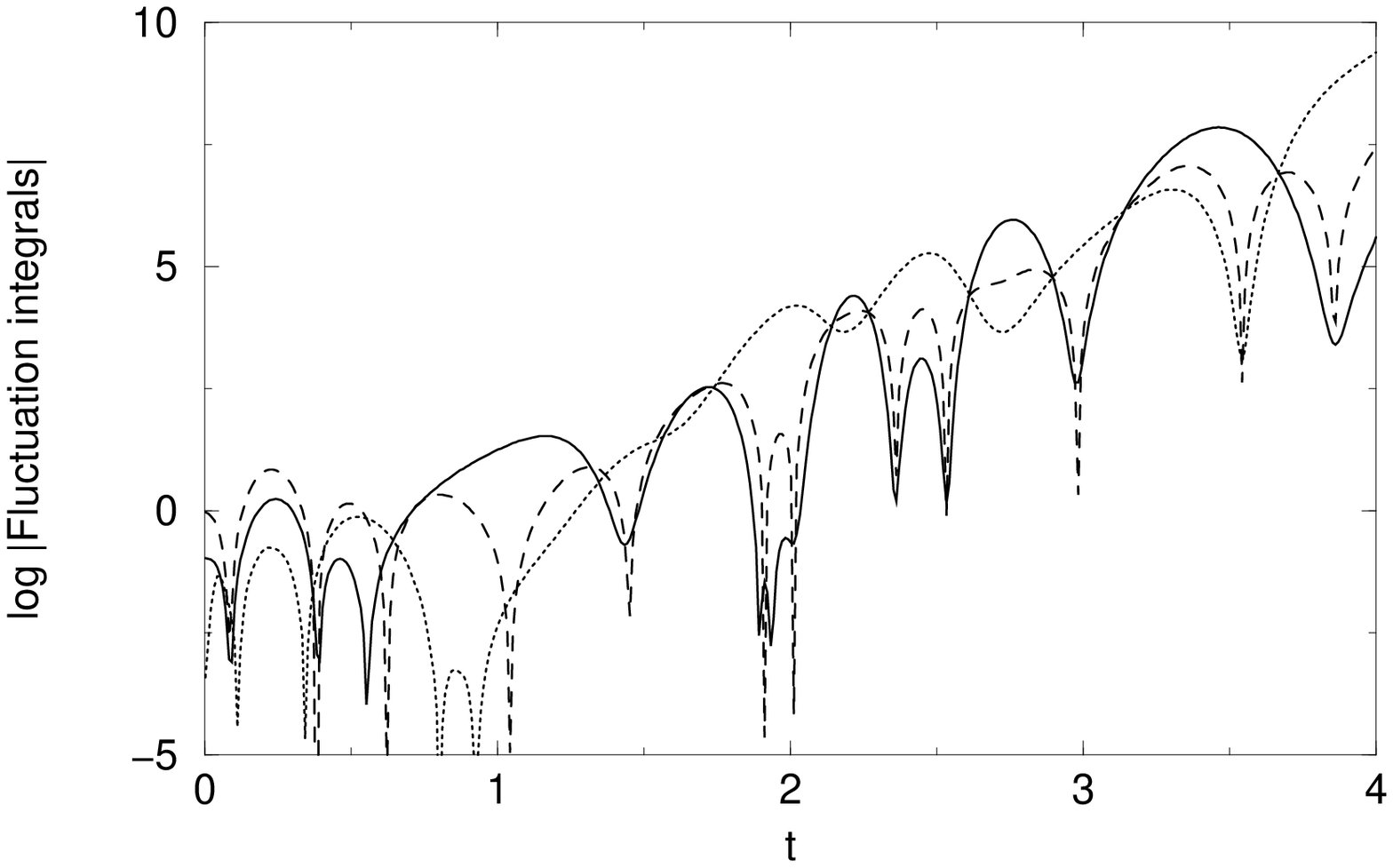}}
\end{center}}
\parbox[t]{8cm}
{{\small FIG. 1:  
The log of the fluctuation integrals 
$\langle X_1^2 \rangle$ (solid), 
$\langle X_1 X_2 \rangle$ (dashed), 
and $\langle X_2^2 \rangle$ (dotted) for the narrow resonance regime.
}}
\hspace{0.5cm}\parbox[t]{8cm}
{{\small FIG. 2:  
The log of the fluctuation integrals 
$\langle X_1^2 \rangle$ (solid), 
$\langle X_1 X_2 \rangle$ (dashed), 
and $\langle X_2^2 \rangle$ (dotted) for the broad resonance regime.
}}}

\vspace{0.5cm}

Fig.~1 shows the log of the three fluctuation integrals $\langle X_1^2
\rangle$, $\langle X_1 X_2 \rangle$ and $\langle X_2^2 \rangle$ for
the narrow resonance case.  These are seen to be dominated by the
exponential growth of $\langle X_2^2 \rangle$, while the other
contributions grow more slowly. Therefore, the evolution is
characterized by production of $X_2$ particles.  We now turn to the
broad resonance regime, where things look quite different. Here, each
of the fluctuation integrals grows rapidly as shown in Fig.~2.
Significant mixing of the species occurs along with copious particle
production.

\vskip7pt

\noindent
\parbox{14.8cm}{
\parbox[t]{8cm}
{\begin{center}
\mbox{\epsfxsize=7.5cm\epsfbox{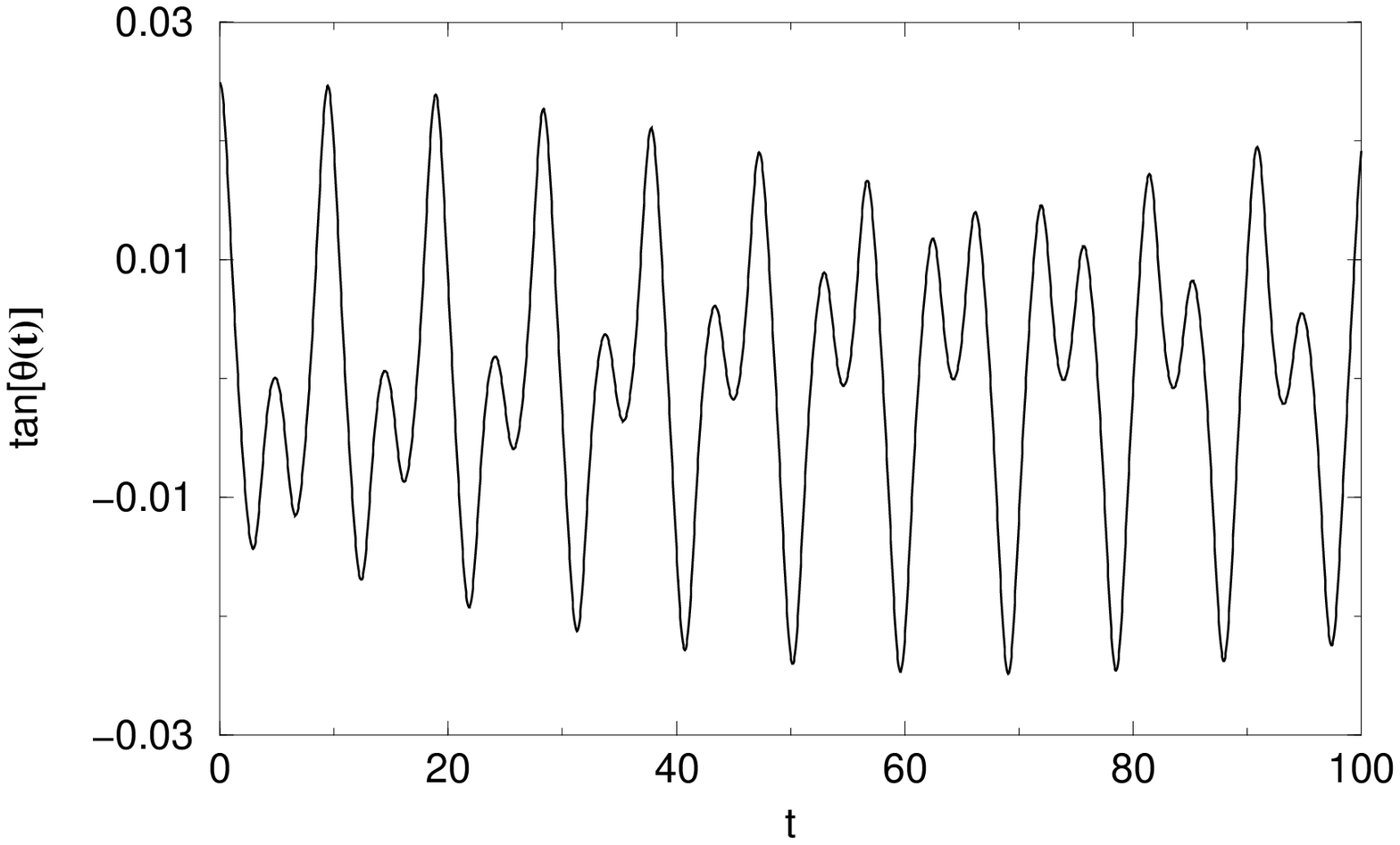}}
\end{center}}
\hspace{.0cm}\parbox[t]{8cm}
{\begin{center}
\mbox{\epsfxsize=7.5cm\epsfbox{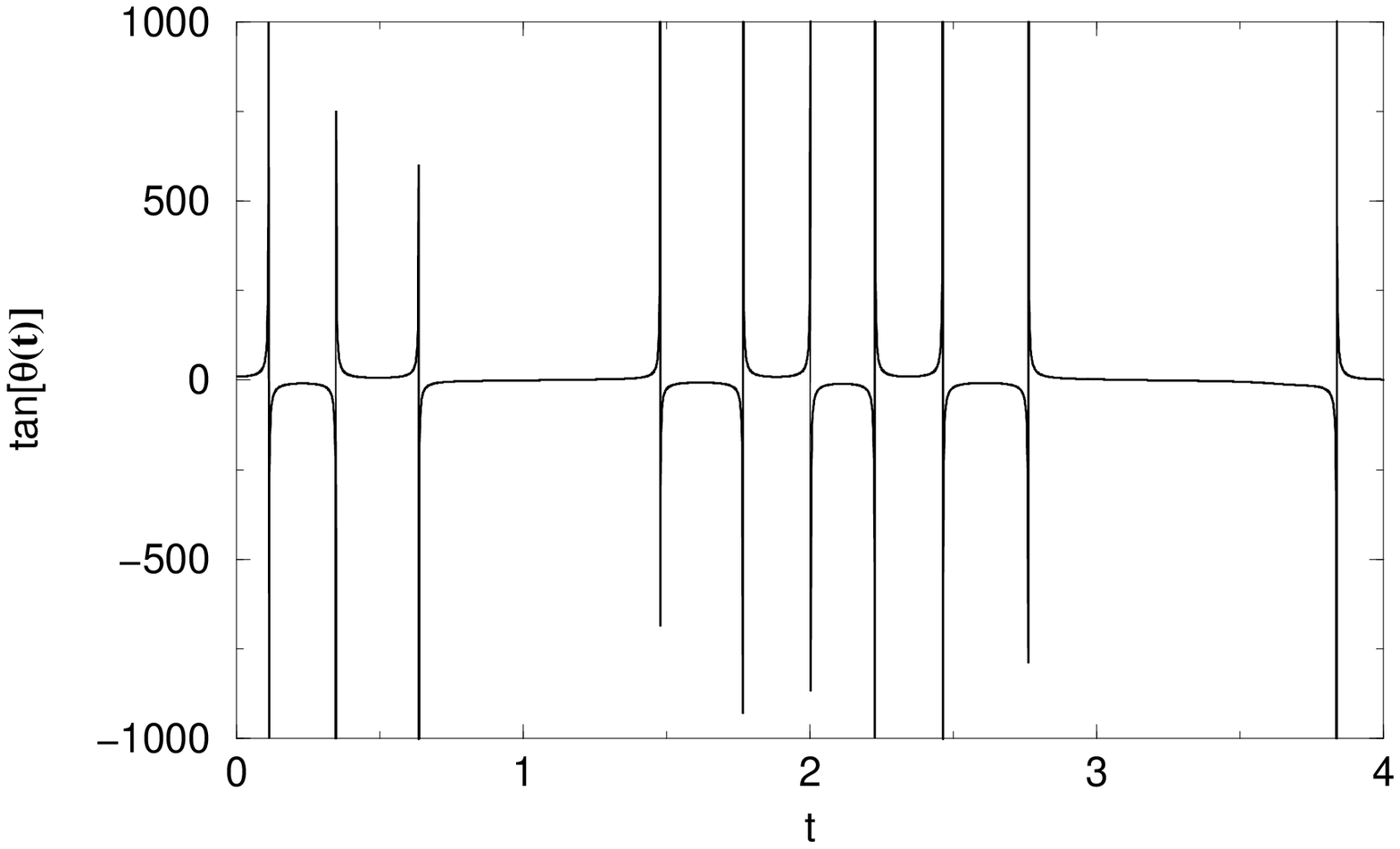}}
\end{center}}
\parbox[t]{8cm}
{{\small
FIG. 3: Time-dependent mixing angle $\Theta (t)$ 
for the narrow resonance regime.}}
\hspace{0.5cm}\parbox[t]{8cm}
{{\small FIG. 4: Time-dependent mixing angle $\Theta(t)$ for the broad
resonance regime.}   
}}

\vspace{0.5cm}

The behavior of the fluctuation integrals is consistent with the
behavior of the time-dependent mixing angle $\Theta$.  Here, the
mixing plays a sub-dominant role in the narrow resonance regime
(Fig.3) with the mixing angle remaining near zero. This means that
$X_2$ predominantly corresponds to the $\chi$ field, such that the
process is one of $\chi$ particle production, which is as expected.
Concentrating on the time-dependent mixing angle in the broad
resonance regime, Fig.~4, significant mixing between the fields is
observed.  The rapid variation in the mixing angle indicates that
mixing between the fields plays a very important role in the evolution
of preheating in the broad resonance regime. The large influence this
has on the behavior of the system is clear from the evolution of the
zero mode components $\phi(t)$ in Fig.~5 and $\chi(t)$ in Fig.~6.

\noindent
\parbox{14.8cm}{
\parbox[t]{8cm}
{\begin{center}
\mbox{\epsfxsize=7.5cm\epsfbox{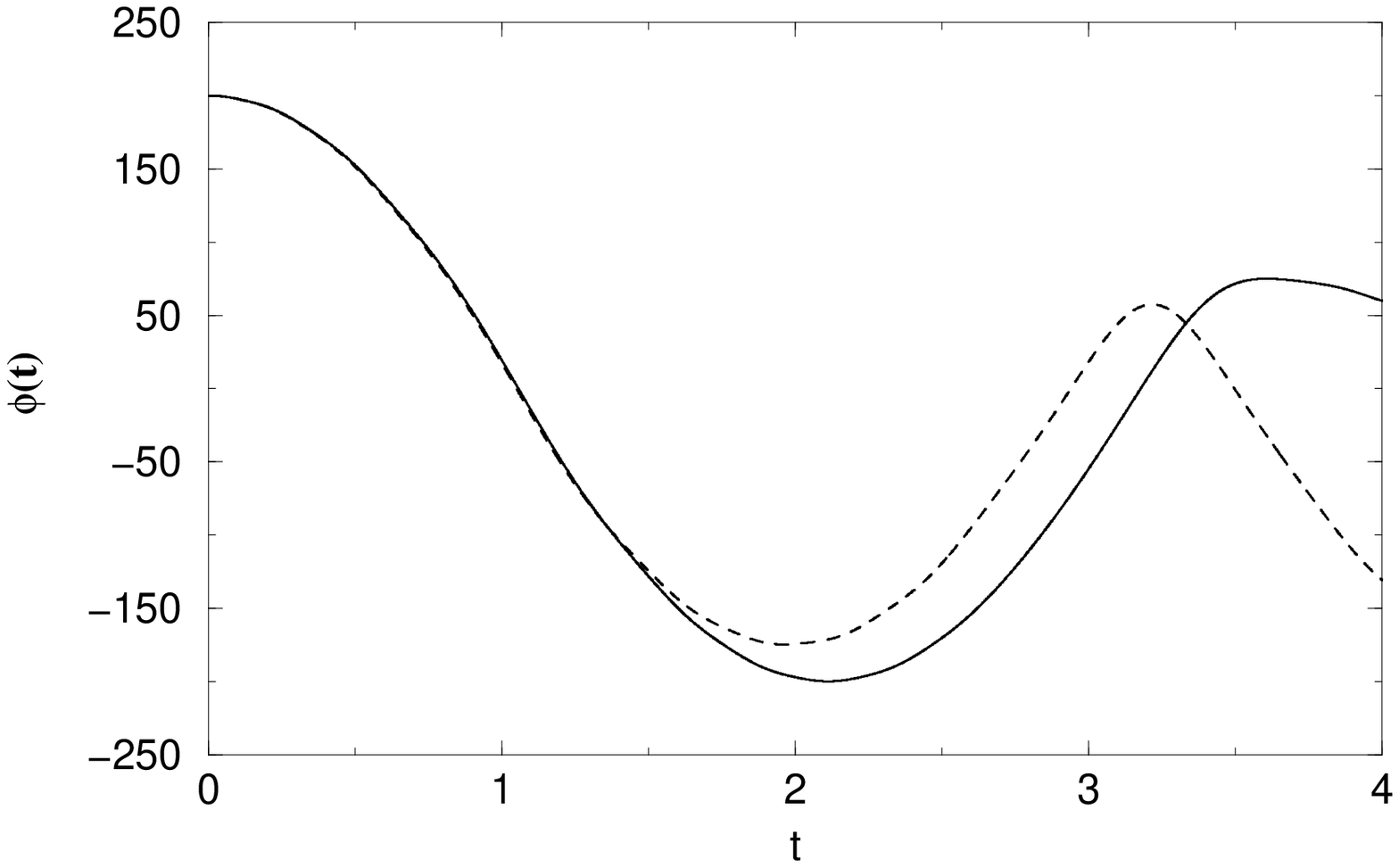}}
\end{center}}
\hspace{.0cm}\parbox[t]{8cm}
{\begin{center}
\mbox{\epsfxsize=7.5cm\epsfbox{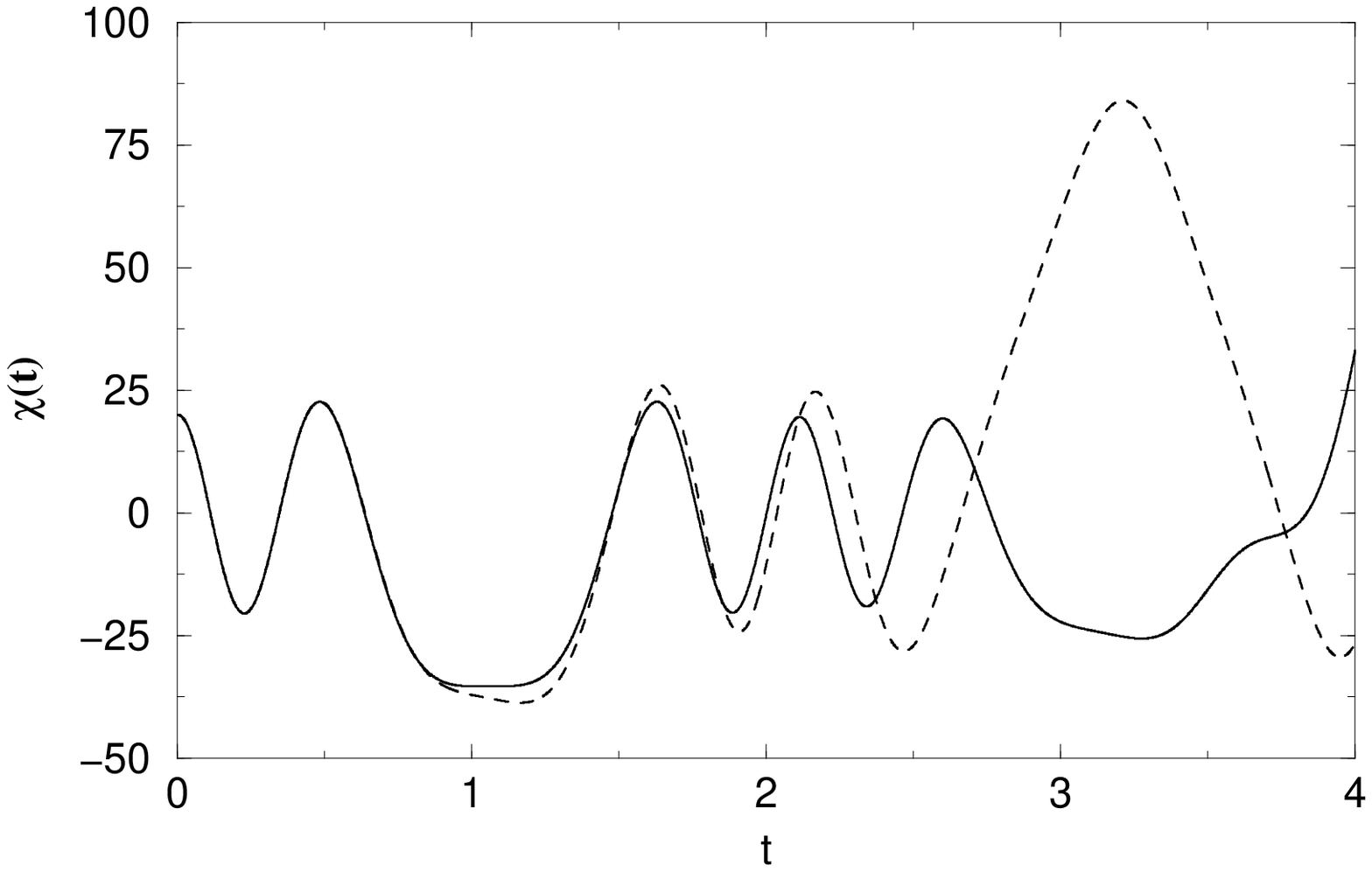}}
\end{center}}
\parbox[t]{8cm}
{{\small
FIG. 5: Zero mode evolution with fluctuation (solid line) 
and without fluctuation
(dotted line) for $\phi(t)$ for the broad resonance regime.}}
\hspace{0.5cm}\parbox[t]{8cm}
{{\small FIG. 6: Zero mode evolution with fluctuation (solid line) 
and without fluctuation
(dotted line) for  $\chi(t)$ for the broad resonance regime.}
}}


\subsection{Supersymmetric Hybrid Inflation}

We now consider a hybrid inflationary model where the finite coupling of 
two fields plays an interesting role in the termination of
slow roll inflation \cite{hybrid}. 
The particular model we study is based on softly broken 
supersymmetry \cite{mar} with the potential
\begin{eqnarray}
\label{pot1}
V(\sigma,N) =\frac{1}{2}m_{\sigma}^2\sigma^2
+\frac{1}{4}\kappa_{\rm s}^2(N^2-2\sigma_c^2)^2
+\kappa_{\rm s}^2N^2\sigma^2\,.
\end{eqnarray}
The field $\sigma$ plays the role of an inflaton during inflation while the 
field $N$ is trapped in a false vacuum $\langle N \rangle =0$. 
The inflaton rolls down the potential
along the $\sigma$ direction to reach a critical value $\sigma =\sigma_c$. Once $\sigma$ 
reaches its critical value, the effective squared mass for the $N$ field 
becomes negative and 
consequently it rolls down from the false vacuum to its global minimum 
through the mechanism of spinodal instability \cite{spinodal,spinodal2}. Thus, 
inflation comes to an end and both the fields begin oscillations around their 
respective minima given by
\begin{eqnarray}
\sigma =0\,, \quad \quad \quad \quad N=\sqrt{2}\sigma_c\,.
\end{eqnarray}

This is the onset of the preheating stage, which has been discussed in
the literature \cite{bellido1,mar,felder}. The difference between the
supersymmetric hybrid potential and non-supersymmetric hybrid
potentials lies in different coupling constants. In Eq.~(\ref{pot1}),
there is only single coupling parameter $\kappa_{\rm s}$, while in the
non-supersymmetric version there can be at least two different
coupling constants. The above potential, except for the $\sigma$ mass
term, can be derived very easily from the superpotential for F-term
spontaneously supersymmetry breaking:
\begin{eqnarray}
W= \frac{\kappa_{\rm s}}{2}\sigma\left(N^2 -\sigma_c^2 \right)\,.
\end{eqnarray}
The appearance of a mass term for $\sigma$ in Eq.~(\ref{pot1}) is due
to the presence of soft supersymmetry breaking. Its presence is
essential for slow roll inflation to produce adequate density
perturbations and also to provide a correct tilt in the power spectrum
\cite{mar}. The height of the potential during inflation is given by
$\kappa_{\rm s}^2 \sigma_c^4$. Similar potentials to Eq.~(\ref{pot1})
can also be derived from D-term supersymmetry breaking as discussed in
Refs.~\cite{susy}. In these models the critical value $\sigma_c$ and
the height of the potential energy are related to the Fayet-Illipoulus
term coming from an anomalous U(1) symmetry.  As in any inflationary
model, hybrid inflation is constrained by COBE \cite{bunn}. This
imposes the bound
\begin{eqnarray}
\kappa_{\rm s} \sigma_c \approx 1.27 \times 10^{15} |\eta|\, \quad {\rm GeV}\,,
\end{eqnarray}
where $\eta$ is one of the slow roll parameters which determines the slope
of the power spectral index \cite{bunn}. For our purpose we fix it to be $|\eta|\sim 0.01$.

\vskip7pt

In order to discuss the details of the physics we mention here the equivalence 
between Eq.~(\ref{pot1}) 
and Eq.~(\ref{pot}). This helps us to establish direct relationship with our 
earlier analysis: 
\begin{eqnarray}
\label{relation}
&\phi& \equiv \sigma\,,\quad \chi \equiv N\,, \quad 
\delta \phi \equiv \delta \sigma\,,\quad \delta \chi \equiv \delta N\,, 
\nonumber \\ 
&\lambda& =0\,, \quad \kappa \equiv 6\kappa_{\rm s}^2\,, \quad
g \equiv 2\kappa_{\rm s}\,, \quad m_{\phi} \equiv m_{\sigma}\,, 
\quad m_{\chi}^2 \equiv -2\kappa_{\rm s}^2\sigma_{c}^2 \,.
\end{eqnarray} 
Notice, that $m_{\chi}^2$ is negative.
An interesting feature of the hybrid model is that irrespective of the
values of the parameters $\kappa_{\rm s}$, $\sigma_c$, and $m_{\sigma}$, as
long as they satisfy
the COBE constraints, the behavior of the mean fields follow a common 
pattern once they begin to oscillate \cite{mar}. First of all, the mass
term for the $\sigma$ field, $m_{\sigma}$, becomes less dominant compared to 
the effective frequency for the two fields, which is given by the effective 
mass for the two fields during oscillations
\begin{eqnarray}
\label{frequency}
m_{{\rm eff}\sigma} = m_{{\rm eff}N} = 2\kappa_{\rm s} \sigma_c \gg m_{\sigma}\,.
\end{eqnarray}
Hence, there is a single natural frequency of oscillation, thanks to
supersymmetry. Since the masses of the fields are the same at the global 
minima, there exists a particular solution of the equations 
of motion for the $\sigma$ and $N$ fields. 
Their trajectory follows a straight line towards their
global minima: 
\begin{eqnarray}
\label{traj}
N =\pm \sqrt{2}\left(\sigma_c -\sigma \right) \,.
\end{eqnarray}

The maximum amplitude attained by the $\sigma$ field is
$\sim\sigma_{c}$, while the other field attains a larger amplitude $N
=\sqrt{2}\sigma_c$. (We remind readers that $\sigma = \sigma_c$
corresponds to the point where the effective mass for $N$ field
changes its sign.) This is the point of instability which we need to
discuss here. From Eq.~(\ref{pot1}), we notice that prior to the
oscillations of the fields, and during the oscillations, the effective
mass square for $\sigma$ is always positive. However, this is not the
case for $N$, and its mass square can be positive as well as negative
even during the oscillations of the fields, provided the amplitudes
are large enough. If the amplitudes for $\sigma$ and $N$ are
such that they satisfy Eq.~(\ref{traj}), then the effective mass
square for the $N$ field is in fact always negative for $\sigma <
\sigma_c$. If the amplitudes are large enough such that after the
second order phase transition the initial amplitude for $\sigma
\approx \sigma_c$, it is then quite possible that near the critical
point the effective mass for the field $N$  vanishes completely.
As far as the motion of the mean field without including fluctuations
is concerned this does not provide any new insight. However, if the
fields are quantized then the perturbations in the field, especially
for $\delta N$, grow exponentially because $\omega^2_{N}$ in
Eq.~(\ref{zerofreq}) becomes negative for sufficiently small momentum
$k$. This shows that the vacuum is unstable near the critical point
$\sigma_c$.

\vskip7pt

Another intuitive way to appreciate this point is to consider the
adiabatic condition for the vacuum. The adiabatic evolution for the
zero mode evolution for $N$ field is given by $|\dot\omega_{N}| \ll
\omega^2_{N}$.  This condition is maximally violated at the point
where the effective mass square for $N$ becomes zero, and, violation
in adiabatic evolution of the zero mode for $N$ suggests that many
fluctuations of $\delta N$ are produced during the finite period when
the adiabaticity is broken \cite{felder}.  This explanation is quite
naive because the overall production of particles and fluctuations
depends also upon the global behavior of the zero mode fields. The
effect of corrections due to fluctuations might affect the production
of particles and this is the point we are going to emphasize in our
numerical simulations.

\vskip7pt

In some sense the hybrid model is quite different from chaotic
inflationary models. In chaotic models, the inflaton field rolls down
with an amplitude $\sim 1/(mt)$, where $m$ is the mass of the
oscillating field. However, in the hybrid model the amplitude of the
oscillations die down very slowly, allowing many oscillations of the
$\sigma$ and $N$ fields in one Hubble time. Thus, one could expect
large amplitude oscillations of the fields for a long time. This
crucially depends on the parameter $\sigma_c$.  If $\sigma_c \ll
M_{\rm p}$, then we notice that effective masses for $\sigma$ and $N$
fields during oscillations are much larger than the Hubble parameter.
The Hubble parameter is given by $H \approx \kappa\sigma_c^2/M_{\rm
p}$ during inflation, so,
\begin{eqnarray}
\frac{m_{{\rm eff}\sigma}}{H} = \frac{m_{{\rm eff}N}}{H} \approx 
\frac{M_{\rm p}}{\sigma_c} \gg 1\,,
\end{eqnarray}
provided the scale of $\sigma_c$ is quite small compared to the Planck
mass, we can effectively neglect the expansion of the Universe.

\vskip7pt

In the supersymmetric hybrid model there are two regimes of
interest. Just after the mass square of the $N$ field becomes
negative, the fields begin to oscillate with an amplitude which
decreases as $\propto 1/t^2$. When the field amplitude drops below
$|N(t)/\sqrt{2}\sigma_c -1|\leq 1/3$, the amplitude of the
oscillations decreases as $\propto 1/t$. In this regime, when the
expansion of the Universe is neglected, the amplitude of the
oscillations remains constant and the oscillations are harmonic:
\begin{eqnarray}
\label{evolv}
\frac{N(t)}{\sqrt{2}\sigma_c}\approx 1 + \frac{1}{3}\cos(m_{{\rm eff}\sigma}t)\,.
\end{eqnarray}
The corresponding evolution equation for the $\sigma$ field can be found
from Eq.~(\ref{traj}). 

\vskip7pt

In this paper we are neglecting the expansion of the Universe. We will 
to concentrate upon two regimes: One with large amplitude oscillations
which leads to the following parameters
\begin{eqnarray} \label{par3}
\lambda&=&0~\kappa=24,~g=4~m_\sigma^2=0,~m_N^2=-16\times 10^{-12},\nonumber\\
\sigma(0)&=&1.4\times 10^{-6},
~N(0)=1\times 10^{-15}\, ,
\end{eqnarray}
and, the other with small amplitude
oscillations with the parameters
\begin{eqnarray} \label{par4}
\lambda&=&0,~\kappa=24\times 10^{-6},
~g=4\times 10^{-3}~m_\sigma^2=0,~m_N^2=-4\times 10^{-12},
\nonumber\\
\sigma(0)&=&0.24\times 10^{-3},
~N(0)=0.66\times 10^{-3}\, .
\end{eqnarray}
The coupling constants are dimensionless while the other dimensionful
parameters denoted in Planck units.  We find below a marked difference
in the zero mode behavior of the fields $\sigma$ and $N$ in these two
cases depending on whether the fluctuations are taken into account or
neglected.

\vskip7pt

In parameter set (\ref{par3}), we study the features of the fields
with a large amplitude. This can happen when the $\sigma$ and $N$ 
fields begin their oscillations just after the end of inflation. As mentioned
earlier,  after the end of inflation the maximum amplitude
attained by the mean fields can be quite large $\sigma = \sigma_c$, and
$N = \sqrt{2}\sigma_c$. This is precisely the initial condition we have chosen 
for the mean fields for our numerics, as shown in Fig.~7. 
The values for $\kappa_s$ and $\sigma_c$ can be evaluated from Eq.~(\ref{relation}),
which yields
\begin{eqnarray}
\kappa_{\rm s} =2\,, \quad \quad \sigma_c = \sqrt{2}\times 10^{-6}\,.
\end{eqnarray}

\noindent
\parbox{14.8cm}{
{\begin{center}
\mbox{\epsfxsize=7.5cm\epsfbox{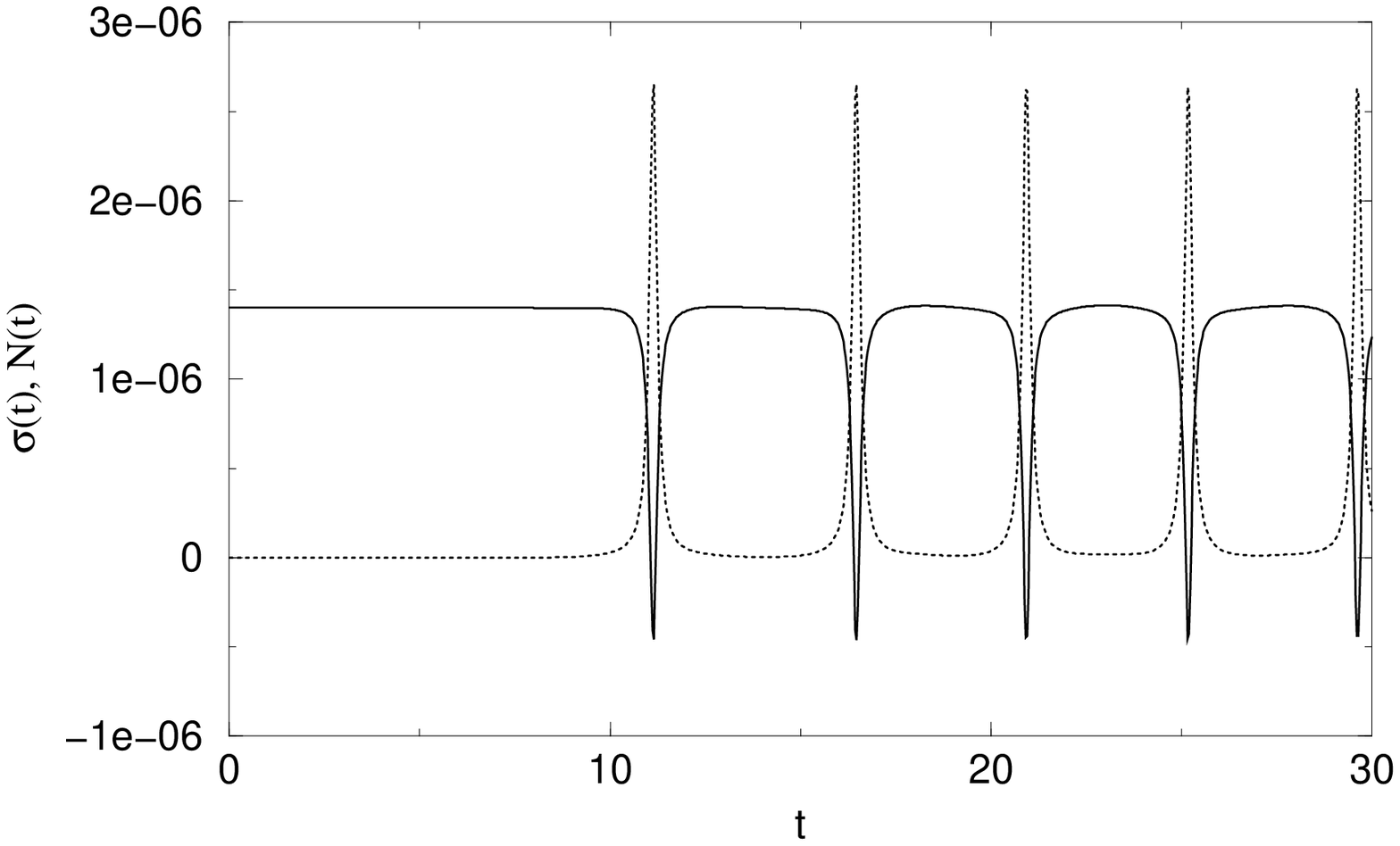}}
\end{center}}
\hspace{.0cm}
{{\small FIG. 7: Evolution of $\sigma$ (solid line) and $N$ (dotted line) without 
fluctuations for parameter set (\ref{par3}). 
}}}

\vspace{.5cm}

\noindent
\parbox{14.8cm}{
\parbox[t]{8cm}
{\begin{center}
\mbox{\epsfxsize=7.5cm\epsfbox{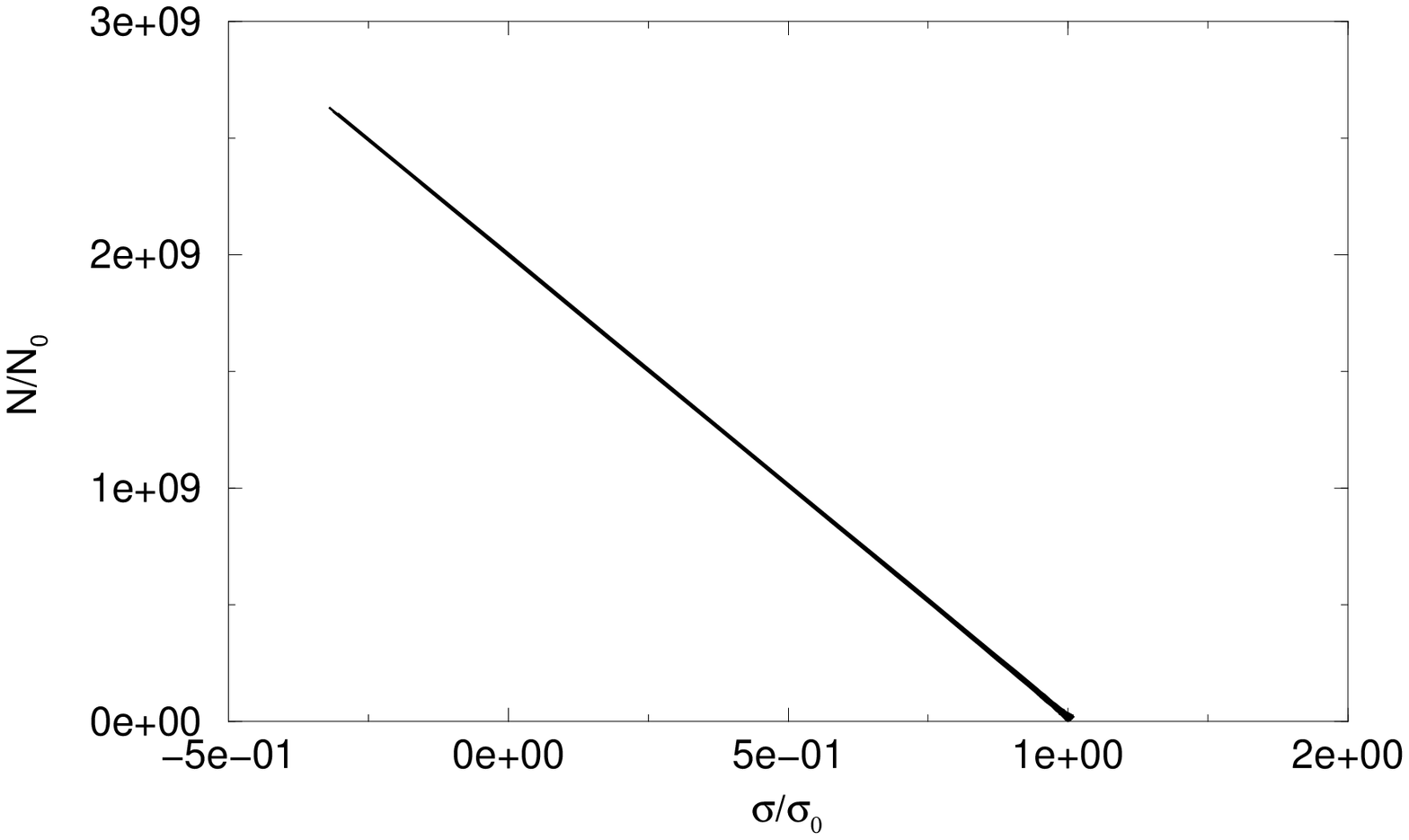}}
\end{center}}
\hspace{.0cm}\parbox[t]{8cm}
{\begin{center}
\mbox{\epsfxsize=7.5cm\epsfbox{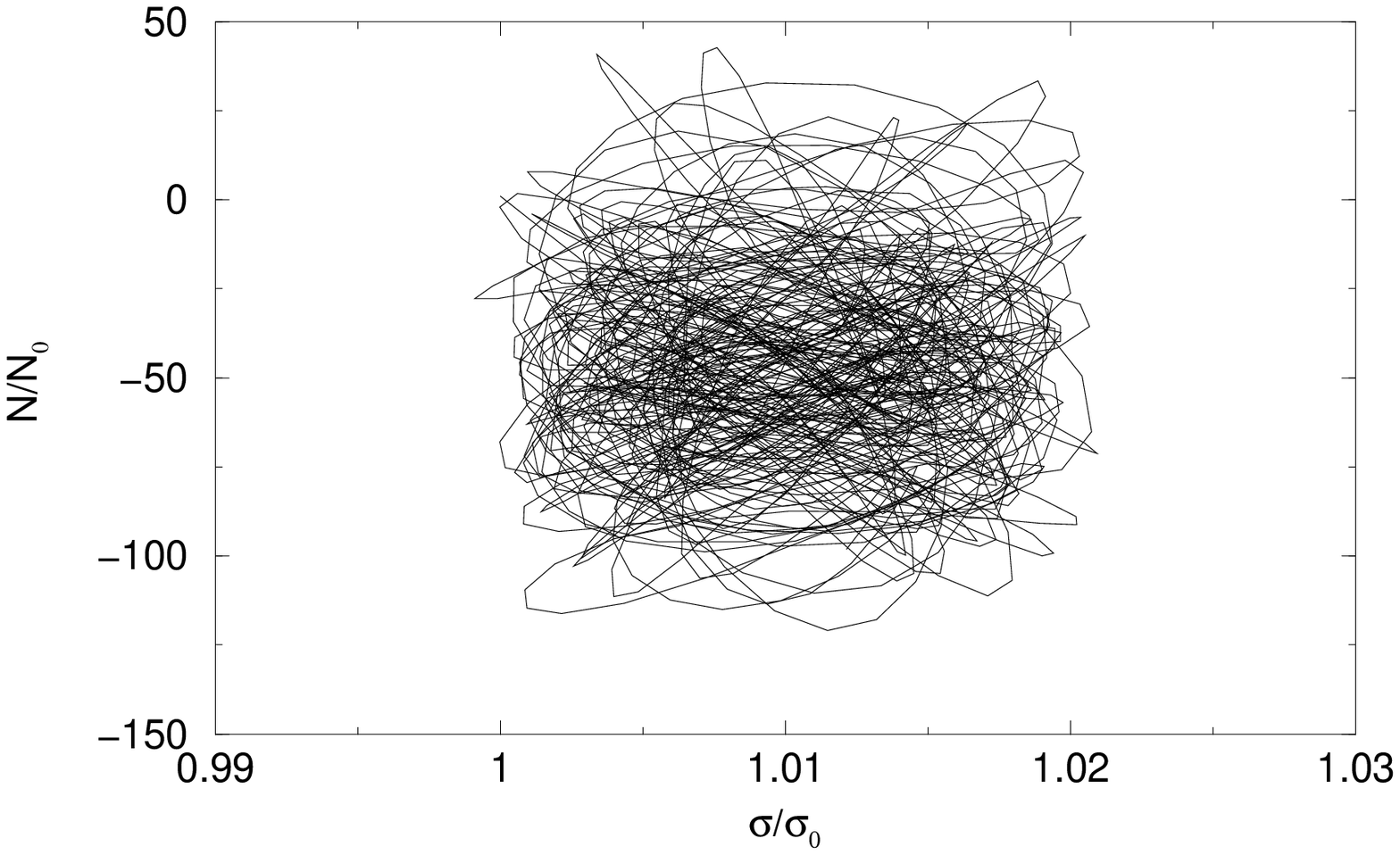}}
\end{center}}
\parbox[t]{8cm}
{{\small
FIG. 8: Trajectory for the fields $\sigma$ and $N$ for 
parameter set (\ref{par3}) without  fluctuations.
}}
\hspace{0.5cm}\parbox[t]{8cm}
{{\small FIG. 9: Trajectory for the fields $\sigma$ and $N$ for 
parameter set (\ref{par3}) with  fluctuations. 
}
}}

\vspace{0.5cm}

We notice that the evolution for $\sigma$ and $N$ 
fields without taking into account the fluctuations are anharmonic, 
see Fig.~7, and, their trajectories
in the $\sigma-N$ plane is a straight line, as shown in Fig.~8. However,
switching on the fluctuations leads to a completely chaotic trajectory as shown in 
Fig.~9. The departure from the straight line trajectory is 
quite significant and it tells us that the renormalized zero mode equations have 
different contributions to the parameter $6\kappa_{\rm s}^2$ and to 
the effective mass of the $N$ field. This mismatch in the frequencies of the
zero mode equations for $\sigma $ and $N$ leads to an irregular trajectory. 

The other way to interpret this
behavior is to think in terms of different effective mass corrections to $\sigma$ and 
$N$ fields, such that the effective frequencies of the oscillations for $\sigma$ and
$N$ do not match each other at the bottom of the potential. This is certainly a
nontrivial result. Nonetheless, the result is quite expected from the fact that the
amplitude of the oscillations are quite large and the effective  
mass for the $N$ field is zero at each and every oscillation
when $\sigma =\sigma_c$.  
  
\noindent
\parbox{14.8cm}{
\parbox[t]{8cm}
{\begin{center}
\mbox{\epsfxsize=7.5cm\epsfbox{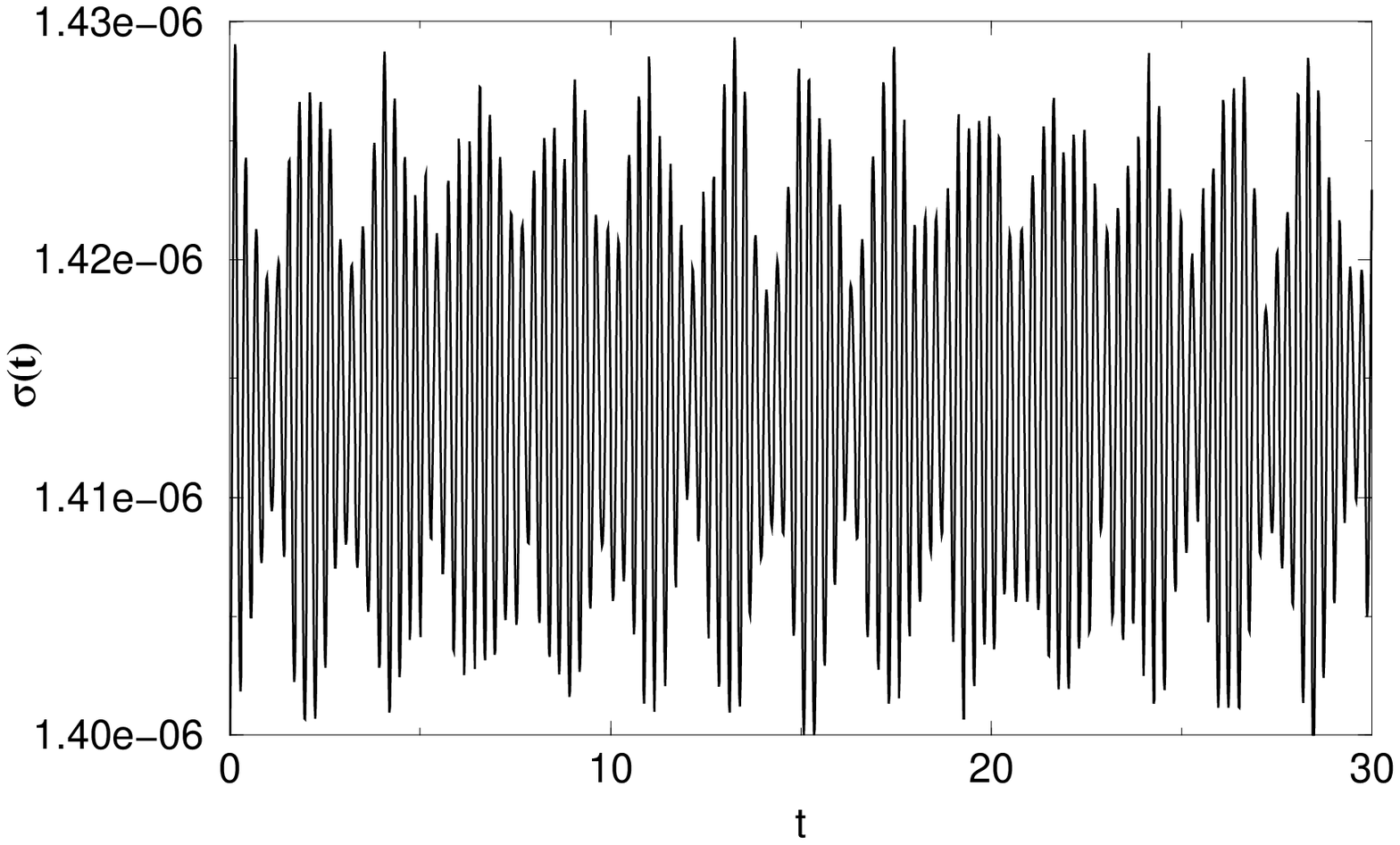}}
\end{center}}
\hspace{.0cm}\parbox[t]{8cm}
{\begin{center}
\mbox{\epsfxsize=7.5cm\epsfbox{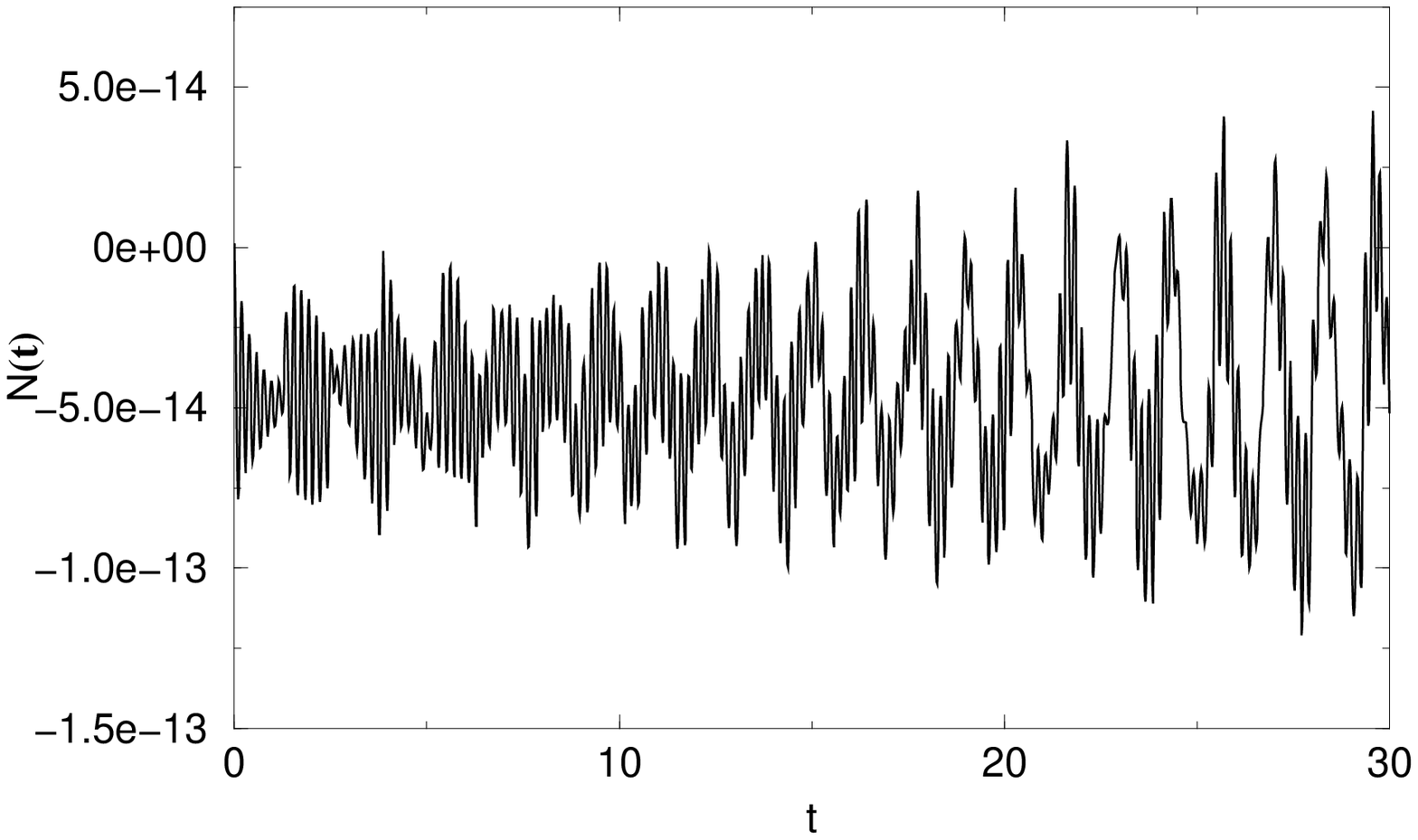}}
\end{center}}
\parbox[t]{8cm}
{{\small
FIG. 10: Zero mode evolution for $\sigma(t)$ with fluctuations
for parameter 
set (\ref{par3}).
}}
\hspace{0.5cm}\parbox[t]{8cm}
{{\small FIG. 11: Zero mode evolution for $N(t)$ with fluctuations
for parameter 
set (\ref{par3}).}
}}

\vspace{0.5cm}

As we mentioned earlier, the frequencies of the oscillations of the
zero modes are different, as can be noticed in Figs.~10 and 11. The
zero mode of $\sigma$ influenced by the fluctuations oscillates around
its minimum $\sigma =0$ with a more rapid frequency than when
fluctuations are neglected.  This suggests that the effective mass
correction to the zero mode for $\sigma$ is  coming solely from the
finite coupling contribution from the $N$ field. (Note, that we have
already set the bare mass for $m_{\sigma} =0$.) The oscillations
maintain the regularity with increasing and decreasing
amplitude. However, the story is not the same for the zero mode
behavior for the $N$ field. The amplitude of $N$ increases gradually
and the frequency of the oscillations varies. We mention here that the
effective mass for the $N$ field can vanish at a critical point. As a
result, the adiabatic condition for the $N$ field is violated at those
instants and this is the reason why the amplitude of the $N$ field is
enhanced rather than suppressed.

\vskip7pt

The evolution of the energy density is shown in Fig.~12. At first
instance it seems quite odd that the energy density of the
fluctuations does not increase further. One would naively expect a
larger contribution of the energy density of $\delta \sigma$ and
$\delta N$. This is not the case here.  The energy density for the
mean fields and the fluctuations are equally shared. The reason is the
correction due to the fluctuations.  These corrections modify the
effective mass of the $N$ field and induce corrections to the coupling
constants, namely $\kappa$ and $g$. The coupling constants are
modified in such a way that the trajectory of zero mode fields become
irregular.  Usually the production of fluctuations is not efficient in
this case. This is quite similar to the situation of preheating in
non-supersymmetric hybrid models \cite{bellido1}. Even though we
started with a supersymmetric hybrid model where at the bottom of the
potential there is a single effective frequency, the situation changes
completely if the fluctuations are taken into account. Essentially the
coupling constants get a large correction which does not preserve an
effective single coupling constant for the evolution of the zero mode
fields.  This is precisely the reason why the zero mode trajectory
becomes irregular and also why the production of $\delta \phi$ and
$\delta N$ is so low.

\noindent
\parbox{14.8cm}{
{\begin{center}
\mbox{\epsfxsize=6.5cm\epsfbox{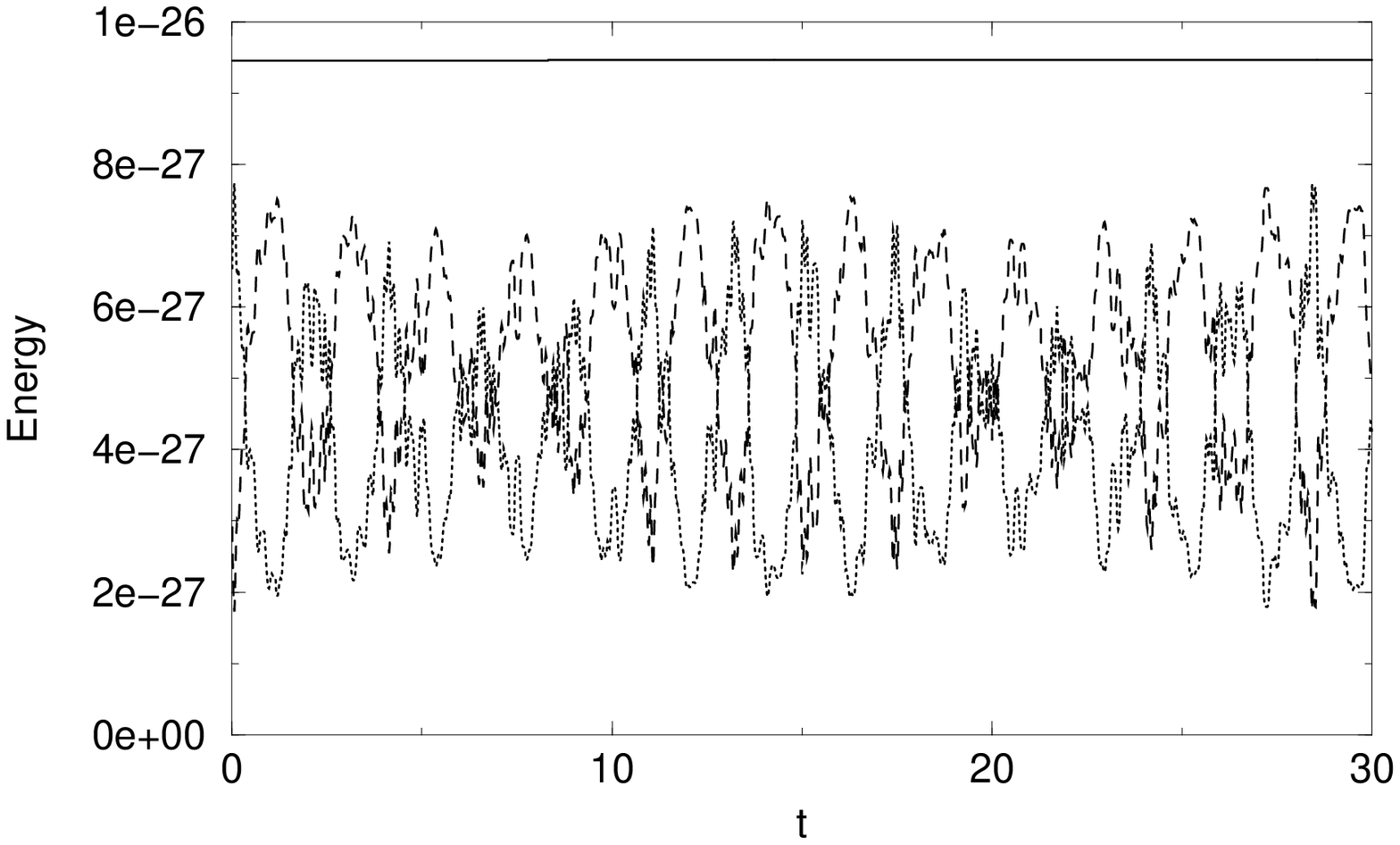}}
\end{center}}
{{\small FIG. 12: Energy density stored in $\delta \sigma$ and $\delta N$ fluctuations
(dashed line) and 
zero mode energy density (dotted line) for parameter set (\ref{par3}).
The total energy remains constant (solid line).
}}
}

\vspace{0.5cm}

As a next example of the supersymmetric hybrid model we choose parameter set 
(\ref{par4}), with a small coupling $\kappa_{\rm s}$ and small $\sigma_c$ 
\begin{eqnarray}
\kappa_{\rm s}=2\times 10^{-3}\,, \quad \quad \sigma_c = 0.707 \times 10^{-3}\,.
\end{eqnarray}
In this example, the coupling between the fields is quite small; $g
=4\times 10^{-3}$, and also the initial conditions for $\sigma$ and
$N$ have been chosen such that the fields oscillate around their
respective minima. The maximum amplitude for $\sigma(0) = \sigma_c/3$
and $N(0) = (2\sqrt{2}/3)\sigma_c$ is much below the critical point
$\sigma_c$. We remind the readers that the chosen initial conditions
for the oscillations do not come naturally just after the end of
inflation, therefore this example does not represent a real
situation. In spite of this we study the particular situation in order
to notice the contrast in the behavior of the zero modes and the
energy densities in the fluctuations. This particular set of initial
conditions for $\sigma(0),N(0)$ offers an alternative example where
spinodal instability in the $N$ field does not take place. As a result
the effective mass square for the $N$ field never crosses zero and the
adiabatic condition for the $N$ field is not strongly broken. The
oscillations of the mean fields $\sigma$ and $N$ are harmonic in
nature, as shown in Figs.~13 and 14 by the dotted lines. The
amplitudes are constant with a frequency given by
Eq.~(\ref{frequency}). The oscillations of the mean fields is governed
by Eqs.~(\ref{traj}) and (\ref{evolv}). The trajectory in the
$\sigma--N$ plane is a straight line whose slope is governed by
Eq.~(\ref{traj}).

\vskip7pt

The effect of the fluctuations is also quite expected in this
case. The amplitudes of the zero mode for $\sigma$ and $N$ fields
decreases after a while and, in contrast to the preceding example, the
frequency of the oscillations do not change very dramatically; see the
behavior of zero mode in solid lines in Figs.~13 and 14. The
trajectories for the zero mode evolution remain a straight line in
this case as shown in Fig.~15. This is quite reasonable for the
parameters we have chosen, but an important observation is that the
effect of fluctuations does not alter the straight line trajectory for
the zero mode fields. This suggests that for small amplitude
oscillations the corrections to the coupling constants; $\Delta
\kappa$ and $\Delta g$ are such that the zero mode equations still
have a similar oscillating frequency. This can be seen in Figs.~13 and
14.  The production of $\delta \sigma$ and $\delta N$ is not very
significant because the energy density stored in $\delta \sigma$ and
$\delta N$ does not grow rapidly. Thus the energy transfer from the
zero modes to the fluctuation modes is not  favorable for such a
small amplitude oscillations as can be seen in Fig.~16.

\noindent
\parbox{14.8cm}{
\parbox[t]{8cm}
{\begin{center}
\mbox{\epsfxsize=6.5cm\epsfbox{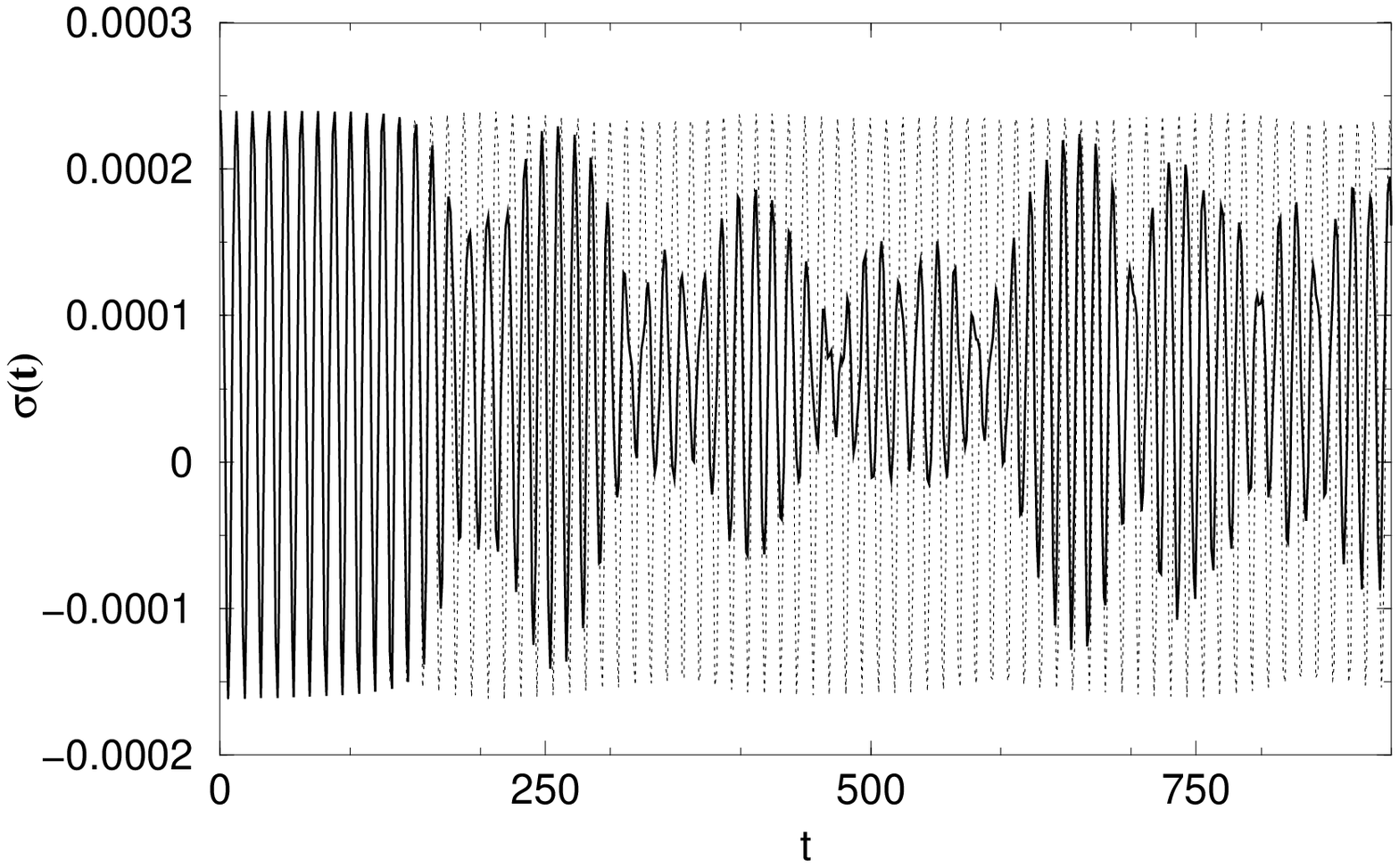}}
\end{center}}
\hspace{.0cm}\parbox[t]{8cm}
{\begin{center}
\mbox{\epsfxsize=7.5cm\epsfbox{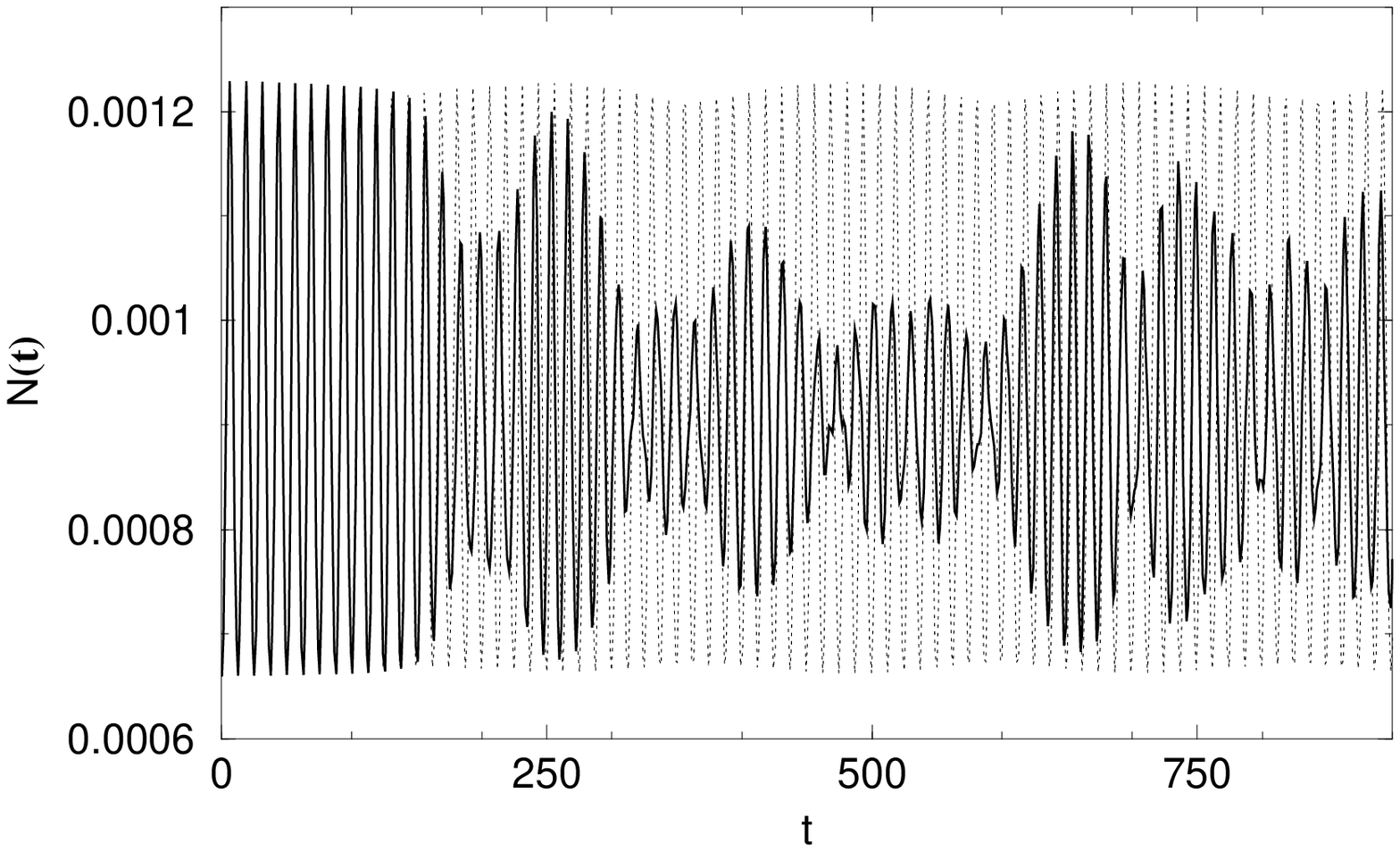}}
\end{center}}
\parbox[t]{8cm}
{{\small FIG. 13:  Zero mode evolution for $\sigma(t)$  with fluctuations (solid lines)
 and the  
 without fluctuations (dotted line) for parameter set (\ref{par4}).
}}
\hspace{0.5cm}\parbox[t]{8cm}
{{\small FIG. 14: Zero mode evolution for $\chi(t)$  with fluctuations (solid lines)
 and without fluctuations (dotted line) for parameter set (\ref{par4}).}
}}


\noindent
\parbox{14.8cm}{
\parbox[t]{8cm}
{\begin{center}
\mbox{\epsfxsize=7.5cm\epsfbox{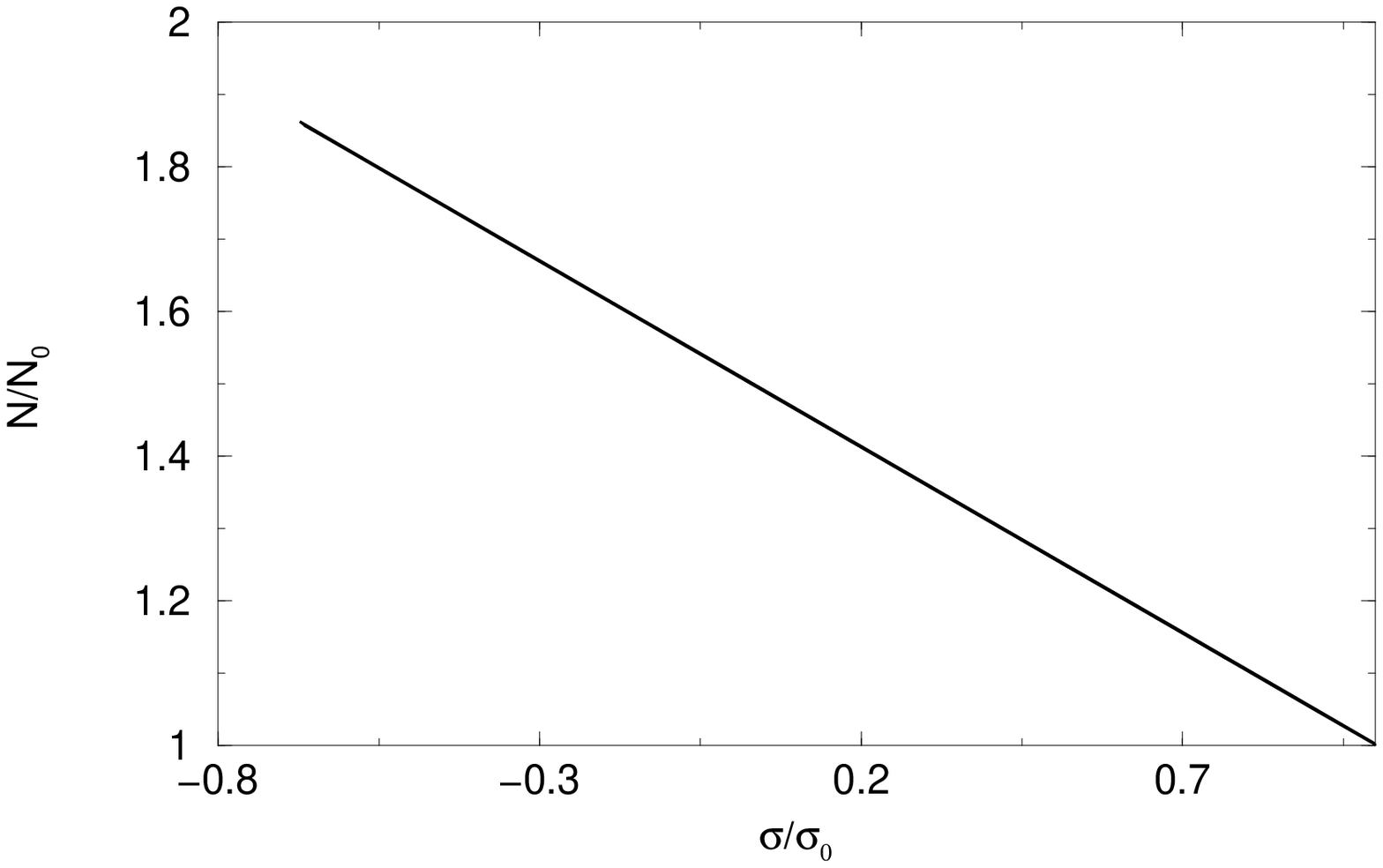}}
\end{center}}
\hspace{.0cm}\parbox[t]{8cm}
{\begin{center}
\mbox{\epsfxsize=7.5cm\epsfbox{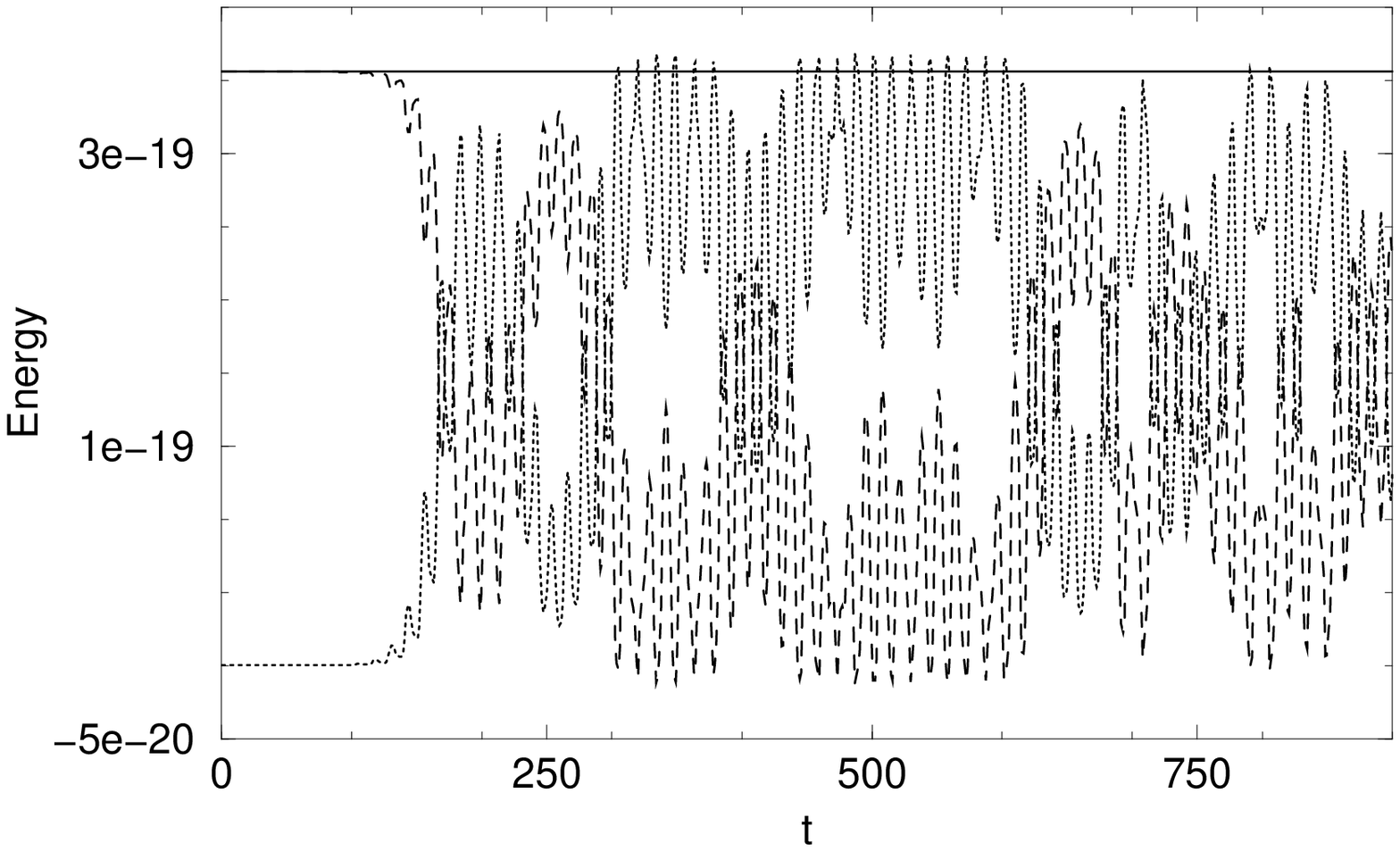}}
\end{center}}
\parbox[t]{8cm}
{{\small FIG. 15:  
Zero mode trajectories with fluctuations for parameter set 
(\ref{par4}).
}}
\hspace{0.5cm}\parbox[t]{8cm}
{{\small FIG. 16: 
Energy density stored in  $\delta \sigma$ and $\delta N$   
(dotted line) along with the zero mode energy density (double dotted line) 
for parameter set 4. The total energy density remains constant (solid line).
}
}}

\vspace{0.5cm}

We conclude this section by mentioning that preheating in this
supersymmetric hybrid model is quite interesting. Depending on the
amplitude of the oscillations of the fields, the behavior of the zero
mode can be quite different.  As a new feature we noticed that if the
amplitude of the oscillations is close to the critical value
$\sigma_c$, the effective mass square for the $N$ field becomes
negative and as a result the fluctuations of the field grows
exponentially.  However, the effect of fluctuations alters the
coupling constants in such a way that the trajectory of the zero modes
become irregular. Even though the adiabatic conditions seem to be
broken for the $N$ field near the critical value, the energy density
transferred from the zero mode to the fluctuations is not
sufficient. Our study reveals
some interesting messages which we briefly mention here. We emphasize
the point that the departure from the straight line trajectory of the
zero mode is an essential feature of a supersymmetric hybrid model if
the fluctuations are taken into account.  Even though, we have not
included the Hubble expansion, the results we have obtained are quite
robust because supersymmetric hybrid inflationary models have a unique
behavior of the fields which allows a smaller inflationary scale
compared to the effective masses of the fields around their global
minima. This suggests that during the oscillations, the expansion is
felt much later, on a time scale determined by the parameters. This
behavior is not shared by models where inflation is governed by a
single field as in chaotic inflationary models. This undermines the
production of quanta from the vacuum fluctuations. In several ways
this affects the post inflationary radiation era of the
Universe. Supersymmetric, weakly interacting dark matter formation and
generation of baryonic asymmetry in the Universe during preheating are
the two most important frontiers which due to our results may warrant
a careful revaluation.

\vskip7pt

In order to substantiate our claim that a due consideration of fluctuations
 after the end of inflation is an important feature of any
supersymmetric hybrid model, we have chosen an unphysical example
which serves the purpose of making a vivid distinction. We stress here
that the spinodal instability which is actually responsible for
producing an irregular trajectory of the zero mode of the fields in a
phase space is completely lacking if the amplitudes of the
oscillations for $\sigma,N$ are small compared to the critical value
$\sigma_c$.  This acts as a comparative study and shows that after the
end of inflation, in a supersymmetric hybrid inflationary model, due to
the spinodal instability in a field, a proper renormalization of the
masses and the coupling constant have to be taken into account.

\section{Conclusion}

We have introduced a formalism to address the dynamics of $N$
nonequilibrium, coupled, time varying scalar fields. We have shown
that the one-loop corrections to the mean field evolution can be
renormalized by dimensional regularization.  For the sake of clarity
and simplicity we restricted ourselves to Minkowski spacetime while
deriving the renormalized equations of motion and the energy density
of the system. We applied our formalism to a two field case where we
study the behavior of the quantized mode functions and the effect of
fluctuations on the zero mode equations of motion for various
parameters including small and large amplitude oscillations and large
and weak coupling between two scalar fields. The varied couplings and
amplitudes illustrate various facets of the intertwined dynamics of
the two fields which lead to a deeper understanding of the production
of self quanta and transfer of energy density between the fields in a
cosmological context.

\vskip7pt

As a special example we have chosen a two field inflationary model
which is genuinely motivated by supersymmetry and thus preserves the
effective masses of the fields to be the same in their local
minima. The model, as a paradigm, predicts inflation which comes to an
end via a smooth phase transition, and robustness of the model is
confirmed by a slightly tilted spectrum of scalar density
fluctuations within the COBE limit.  The model parameters can be
adjusted to give an inflationary scale covering a wide range of energy
scales from TeV to $10^{15}$~GeV. The phase transition leads to a
spinodal instability in one of the fields which leads to a coherent
oscillations of the fields around their global minima. The instability
occurs in one of the fields which demands careful study of the back
reaction to an otherwise growing mean field in an intertwined coupled
bosonic system. An account of influence of the fluctuations gives
rise to uneven contribution to the renormalized masses of the
fields. This results in an irregular trajectory of the zero mode in a
phase space, which breaks the coherent oscillations of the two
fields. This prohibits an excessive production of particles from the
vacuum fluctuations. This requires a careful revaluation of the
successes of the production of weakly interacting massive particles
and baryogenesis via out of equilibrium decay in supersymmetric
hybrid inflationary models. Our study implies that exciting higher
spin particles from the vacuum fluctuations of the coherent
oscillations of the fields in a supersymmetric hybrid inflationary
model demands careful reconsideration.

\vskip7pt  

Even though, we have neglected the effect of expansion in our
calculation, our results are robust enough to claim that the
fluctuations in a supersymmetric hybrid model do not grow if the back 
reaction of the fluctuations are taken into account in the mean field
evolution. An extension of our formalism to an expanding Universe
deserves separate attention.

\section{Acknowledgements}
The authors are thankful to Mar Bastero-Gil and Michael G. Schmidt 
for helpful discussion. We thank Salman Habib for helpful comments on the manuscript. 
A.M. is partially supported by 
{\bf The Early Universe Network} HPRN-CT-2000-00152.
 

\end{document}